\documentclass[aps,pra,letterpaper,twocolumn,showpacs,preprintnumbers,amsmath,amssymb,superscriptaddress]{revtex4}

\usepackage{graphicx}
\usepackage{dcolumn}
\usepackage{bm}
\usepackage{hyperref}

\usepackage{color}

\usepackage{CJK}

\begin{document}

\begin{CJK*}{UTF8}{gbsn}

\preprint{draft version: \today}

\title{High Resolution Spectroscopy of Neutral Yb Atoms in a Solid Ne Matrix}

\author{R. Lambo}
\thanks{These two authors contributed equally.} 
\affiliation{Shenzhen Institutes of Advanced Technology, Chinese Academy of Sciences, Shenzhen, 518055, China}
\author{C.-Y. Xu (徐晨昱)}
\thanks{These two authors contributed equally.} 
\affiliation{Physics Division, Argonne National Laboratory, Argonne, Illinois 60439, USA}
\affiliation{Department of Physics and Enrico Fermi Institute, University of Chicago, Chicago, Illinois 60637, USA}
\author{S. T. Pratt}
\affiliation{Chemical Science and Engineering Division, Argonne National Laboratory, Lemont, Illinois 60439, USA}
\author{H. Xu (徐红)}
\affiliation{Chemical Science and Engineering Division, Argonne National Laboratory, Lemont, Illinois 60439, USA}
\author{J. C. Zappala}
\affiliation{Physics Division, Argonne National Laboratory, Argonne, Illinois 60439, USA}
\author{K. G. Bailey}
\affiliation{Physics Division, Argonne National Laboratory, Argonne, Illinois 60439, USA}
\author{Z.-T. Lu (卢征天)}
\affiliation{Hefei National Laboratory for Physical Sciences at the Microscale, CAS Center for Excellence in Quantum Information and Quantum Physics, University of Science and Technology of China, 96 Jinzhai Road, Hefei 230026, China}
\author{P. Mueller}
\affiliation{Physics Division, Argonne National Laboratory, Argonne, Illinois 60439, USA}
\author{T. P. O'Connor}
\affiliation{Physics Division, Argonne National Laboratory, Argonne, Illinois 60439, USA}
\author{B. B. Kamorzin}
\affiliation{CEST, Skolkovo Institute of Science and Technology, Skolkovo Innovation Center, Moscow  121205,  Russia}
\author{D. S. Bezrukov}
\affiliation{CEST, Skolkovo Institute of Science and Technology, Skolkovo Innovation Center, Moscow  121205,  Russia}
\affiliation{Department of Chemistry, M.V. Lomonosov Moscow State University, Moscow 119991, Russia}
\author{Y. Xie}
\affiliation{Shenzhen Institutes of Advanced Technology, Chinese Academy of Sciences, Shenzhen, 518055, China}
\author{A. A. Buchachenko}
\email{a.buchachenko@skoltech.ru}
\affiliation{CEST, Skolkovo Institute of Science and Technology, Skolkovo Innovation Center, Moscow  121205,  Russia}
\author{J. T. Singh}
\email{singhj@frib.msu.edu}
\affiliation{Physics Division, Argonne National Laboratory, Argonne, Illinois 60439, USA}
\affiliation{National Superconducting Cyclotron Laboratory, Michigan State University, East Lansing, Michigan 48824, USA}



\date{\today}

\begin{abstract}
We present an experimental and theoretical study of the absorption and emission spectra of Yb atoms in a solid Ne matrix at a resolution of 0.025 nm.
Five absorption bands were identified as due to transitions from the $4f^{14}5d^06s^2\ ^1\!S_0$ ground state configuration to $4f^{14}5d^06s6p$ and $4f^{13}5d^16s^2$ configurations. The two lowest energy bands were assigned to outer-shell transitions to $6s6p\ ^3P_1$ and $^1P_1$ atomic states and displayed the structure of a broad doublet and an asymmetric triplet, respectively. The remaining three higher-frequency bands were assigned to inner-shell transitions to distinct $J=1$ states arising from the $4f^{13}5d^16s^2$ configuration and were highly structured with narrow linewidths. A classical simulation was performed to identify the stability and symmetry of possible trapping sites in the Ne crystal.
It showed that the overarching 1+2 structure of the high frequency bands could be predominantly ascribed to crystal field splitting in the axial field of a 10-atom vacancy of $C_{4v}$ symmetry. 
Their prominent substructures were shown to be manifestations of phonon sidebands associated with the zero-phonon lines on each crystal field state. Unprecedented for a metal-rare gas system, resolution of individual phonon states on an allowed electronic transition was possible under excitation spectroscopy which reflects the semi-quantum nature of solid Ne. In contrast to the absorption spectra, emission spectra produced by steady-state excitation into the $^1P_1$ absorption band consisted of simple, unstructured fluorescence bands.


\end{abstract}


\maketitle

\end{CJK*}

\section{Introduction}

\begin{figure}[t]
\centering
\includegraphics[width=0.35\textwidth]{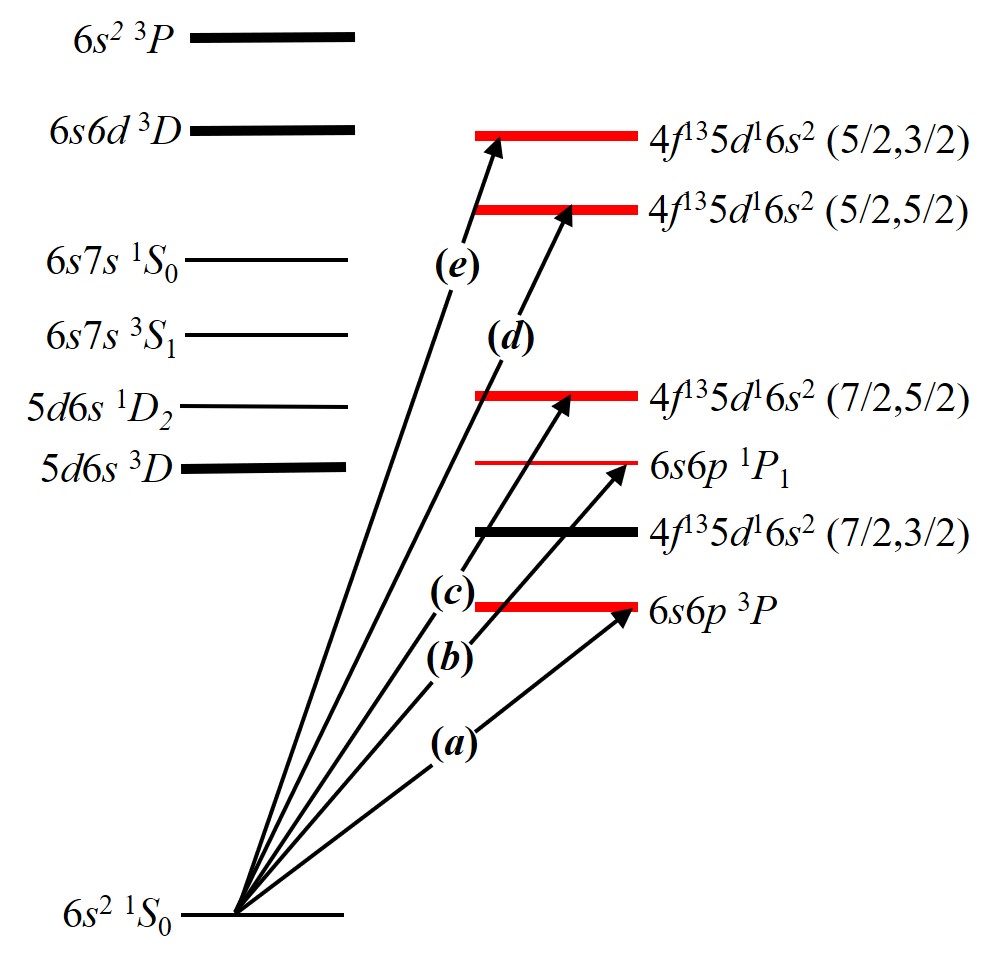}
\caption{Low-lying energy levels of the Yb atom with the fine structures omitted.
The arrows represent five transitions observed in solid Ne labelled from (a)-(e) following Tables \ref{tab:abs} and \ref{tab:emis}. They correspond to $E1$ transitions from the $6s^2\ ^1\!S_0$ ground state to $J=1$ states in five fine structure manifolds.
\label{fig:levels}}
\end{figure}

Rare gas (RG) solids, formed at cryogenic temperatures, are a promising host medium for the capture, detection and quantum state manipulation of guest atoms and molecules.
They provide stable and chemically inert isolation and confinement for a wide variety of guest species at a tunable density---from a single isolated atom to a number possibly exceeding $\sim10^{16}$ atoms/cm$^3$.
Because RG solids are transparent at optical wavelengths, the guest species can be probed using lasers and the induced fluorescence efficiently detected outside the solid.
Spin coherence times of the guest species, which are ultimately dominated by long-range dipolar couplings \cite{vv48,ka53}, could be made as long as $10^3$ s for nuclear spins and $1$ s for electronic spins by minimizing spin impurities within the host matrix.
Applications of this ``matrix isolation'' technique include tests of fundamental symmetries \cite{kd06,edm3}, magnetometry \cite{kanagin13,upadhyay20} and quantum information science \cite{wolf21}. 

Examples of systems studied specifically for these applications include alkali atoms in both solid RGs \cite{wp65} and parahydrogen \cite{upadhyay19a}, Yb atoms in solid Ne \cite{xu11,xu14} and Tm atoms in solid Ar and Kr \cite{gaire19}. For these species, medium effects generally broaden linewidths and shift transition frequencies by several hundred wavenumbers \cite{wp65}. Despite substantial matrix-induced perturbations to their $D$-lines, however, alkali atoms have been successfully optically pumped in matrix isolation \cite{kanagin13, upadhyay16}. Furthermore, spin coherence times as long as 0.1 s have been observed for Rb atoms in solid parahydrogen \cite{upadhyay20}.

Ytterbium (Yb) is a heavy divalent atom with a $^1\!S_0$ electronic ground state, several optical transitions and a naturally abundant isotope, $^{171}$Yb (14\%), whose nuclear spin is 1/2. These features combine to make matrix-isolated $^{171}$Yb a promising candidate for a solid state search for a permanent electric dipole moment \cite{rmp19} or for a nuclear spin-based qubit. Such applications depend on the ability to both optically prepare and readout the $^{171}$Yb nuclear spin state. This in turn depends crucially on the strength of the hyperfine interaction in the excited state compared to the size of the matrix-induced perturbations. The need to characterize the latter motivates the high resolution spectroscopic study of Yb atoms in solid Ne (sNe) presented here. 

Recent work at relatively low resolution ($\sim$ 0.5 nm) has detailed some of the complexities of Yb/RG systems. When solid Xe was used as the matrix host, absorption and emission bands were found to have a two-fold structure due to Yb occupation in tetravacancies and single substitutional sites \cite{kleshhina19Yb}. Earlier work found that when solid Ar was the matrix host, the same bands had a three-fold structure, due to Yb occupation in hexavacancies, tetravacancies and single substitutional sites \cite{Tao2015}. The symmetry of these trapping sites generally induces crystal field splitting (CFS) of degenerate atomic states upon excitation \cite{CrepinTramer}. Low symmetries further multiply the number of bands, while high symmetries create secondary structures due to Jahn-Teller electron-phonon coupling \cite{Bersuker}. The phonon structure itself is usually not visible---except on certain forbidden transitions of the Mn/Kr \cite{by10} and Eu/Ar \cite{br11} systems---and predominantly contributes to the width of strongly non-Frank-Condon bright absorptions. 
 

The above hierarchy of matrix-induced spectral perturbations is typical of the heavy ``classical'' RG solids made of Ar, Kr or Xe. In ``quantum'' matrices, such as those of He and H$_2$, for which the effects of nuclear motion are expected to be much stronger, the identity of distinct trapping sites is eroded and large-amplitude motions magnify electron-nuclear couplings. Solid Ne is often regarded as a ``semi-quantum'' crystal in which one may expect incipient nuclear quantum effects \cite{Cazorla2017}. At the same time, the Ne crystal is a less perturbing environment than the heavier RGs. Isolated in it, the Yb electronic structure is still qualitatively the same as in vacuum, while the lifetimes of excited states are not significantly different \cite{xu11, xu14}.
Based on these observations, single atom detection of Yb in sNe would appear feasible, which could then be extended for use in detecting rare nuclear reactions \cite{loseth19}. This would be in close analogy with the recent demonstration of single atom imaging of Ba in solid Xe, performed with a view to searching for neutrinoless double beta decay\cite{chambers19}.

In order to investigate all these prospects, we have performed the first high resolution (0.025 nm) broadband spectroscopy of the low-lying states of the Yb/Ne system (Fig. \ref{fig:levels}). Five absorption bands were identified as due to Yb atomic transitions from the $^1\!S_0$ ground state of the  $4f^{14}5d^06s^2$ configuration to $J = 1$ states of the $4f^{14}5d^06s6p$ or $4f^{13}5d^16s^2$ configuration. These appear as either broad, largely featureless bands or  narrow, highly structured ones depending on their susceptibility to the matrix-induced perturbations described above. To assist in their interpretation, we performed classical simulations that gave the structure and symmetry of the Yb/Ne trapping sites and whose corresponding theoretical spectra adequately explained the observed lineshapes. The spectrum resulting from steady state $6s^2\ ^1\!S_0 \rightarrow 6s6p\ ^1\!P_1$ excitation was also recorded at the same high resolution. Knowledge of the decay paths originating from the excited singlet state will be important for single atom spectroscopy.

The remainder of this paper is organized as follows. Section \ref{Sec:Experimental} details the experimental methods, while Section \ref{Sec:Results} presents the results. These results, together with the classical simulations, are discussed in Section \ref{Sec:Discussion}. A summary and concluding remarks are made in Section \ref{Sec:Summary}.

\section{Experimental Setup}
\label{Sec:Experimental}

\begin{figure}[t]
\centering
\includegraphics[width=0.45\textwidth]{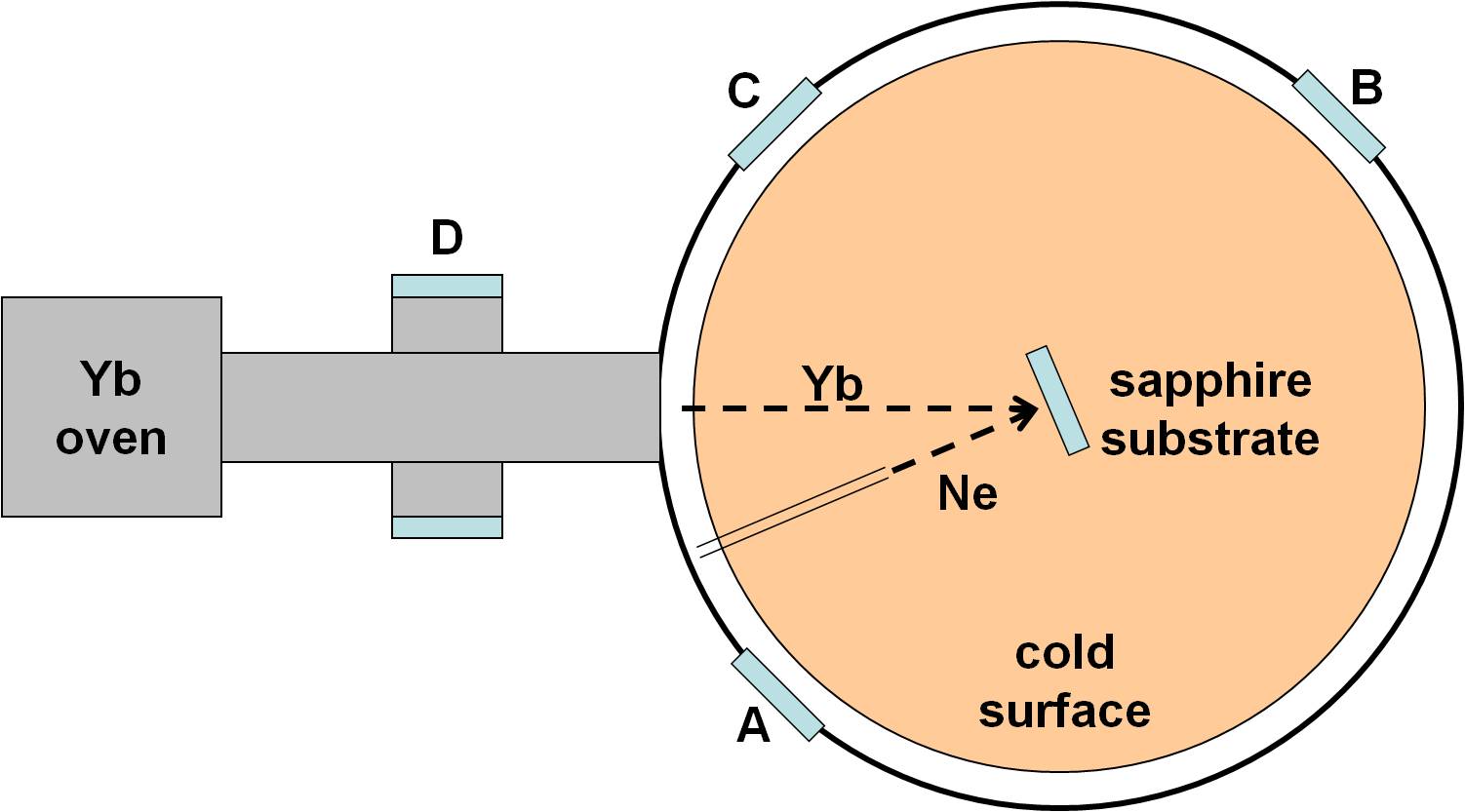}
\caption{A sketch of the cold surface inside the liquid helium cryostat together with the effusion oven and Yb beam line.
A, B and C are three viewports covered with fused silica windows.
D is a six-way cross where the Yb beam intensity is measured.
\label{fig:cryostat}}
\end{figure}

All the experiments were conducted with the liquid helium cryostat illustrated in Fig. \ref{fig:cryostat}. Its cold surface has an area of 300 cm$^2$ that is cooled by the liquid helium bath to 4.2 K, at which temperature the pressure of the vacuum falls below $10^{-8}$ Torr.
The Yb/Ne samples were grown on a 2.54 cm diameter c-plane sapphire substrate installed vertically on the cold surface using a copper mount.
Indium wires and low vapor pressure grease (Apiezon N) were used to achieve good thermal contact between the substrate and the cold surface.
The temperature of the substrate was measured by a resistance temperature detector (Lakeshore CX-1050-AA) and was typically at 4.2-5.0 K depending on the thermal contact.

There were four 2.54 cm diameter ports on the perimeter of the cryostat.
One port was connected to an effusion oven whose crucible was loaded with metallic Yb of natural isotope abundance. The other three ports (A, B and C in Fig. \ref{fig:cryostat}) were closed with fused silica windows and used as viewports.
Viewports A and B, which were each at an angle of 22.5$^\circ$ to the normal direction of the substrate surface, were used during measurements of the sample thickness and the absorption spectroscopy. Viewport C, which was at an angle of 67.5$^\circ$ to the normal direction of the substrate surface, was used in the emission study to collect fluorescence from the sample upon excitation.

The sample was prepared as follows: the Ne gas first flowed through a purifier (LDetek LDP1000) and a 77 K charcoal trap and then leaked into the cryostat via a capillary tube that ended 5 cm away from the substrate. The implantation of Yb atoms was achieved by directing the beam from the effision oven at the substrate while the sNe crystal grew. The sample thickness was determined by recording the transmitted light of a He-Ne laser and counting the fringes due to the thin-film interference effect. Meanwhile, the Yb intensity was measured using the cross-beam-fluorescence method at the six-way cross (viewport D in Fig. \ref{fig:cryostat}). A 398.9 nm laser beam resonant with the Yb $6s^2\ ^1\!S_0 \rightarrow 6s6p\ ^1\!P_1$ transition in vacuum was applied perpendicularly to the Yb beam and the fluorescence measured using a photomultiplier tube along an axis perpendicular to both beams. 

We found empirically that individual Yb atoms (as opposed to higher aggregates) were isolated in sNe only when the Yb:Ne ratio was kept below 5 ppm. In practice, we held that ratio at 1 ppm for which a Yb beam intensity of $4\times10^{14}$ atoms/cm$^2\cdot$hr and an oven temperature of 620 K were required. The steady state Ne growth rate was typically 50 $\mu$m/hr and the Yb areal and volumetric densities were estimated to be $2\times10^{15}$ atoms/cm$^2$ and $10^{16}$ atoms/cm$^3$ respectively. Under these conditions, the heat load due to the deposition of Ne and the blackbody radiation of the oven increased the substrate temperature by only 50 mK. A ramp-up time of two hours was required to preserve the transparency of the sample and, once this rate was reached, it could be grown for five hours before it cracked under its own internal strain.



In the absorption study, a weak broadband light source (Ocean Optics DH2000-DUV) illuminated the sample through viewport A, and the transmitted light was collected through viewport B and fiber-coupled to a spectrometer. In the emission study, a light-emitting diode (LED) centered at 385 nm illuminated the sample through viewport A, and the induced fluorescence was collected at viewport C and also fiber-coupled to a spectrometer. Two spectrometers were available to us to analyze the absorption and the emission signals, both of which used line CCD cameras as the detectors. The Ocean Optics USB4000-UV-VIS spectrometer covered a broad range from 200 nm to 1100 nm and had a resolution of 1.5 nm. The McPherson 225 spectrometer had a resolution of 0.025 nm and covered a range from 0 nm to 600 nm but with a camera frame that was only 40 nm wide.


To obtain the highest possible spectral resolution, we also used an optical parametric oscillator (OPO) (Continuum Sunlite EX) as the probe light.
The OPO was pumped by a third harmonic pulsed Nd:YAG laser (Continuum Powerlite DLS) before passing through a frequency doubler (Inrad Autotracker II) to provide UV light.
Each pulse was about 10 ns long and contained about 1 mJ of energy.
The light was attenuated by four orders of magnitude to avoid melting the sample and damaging the detector.
The output beam was split into a probe beam that went through the sample and a reference beam that traveled outside the cryostat.
The linewidth of the OPO was about 1 GHz and the frequency was scanned in 2 GHz steps to map out the lineshape of the whole absorption band.

\section{Results}
\label{Sec:Results}

\subsection{Probe with Broadband Light}
\label{Sec:Probe_white}

\begin{figure}[t]
\centering
\includegraphics[width=0.48\textwidth]{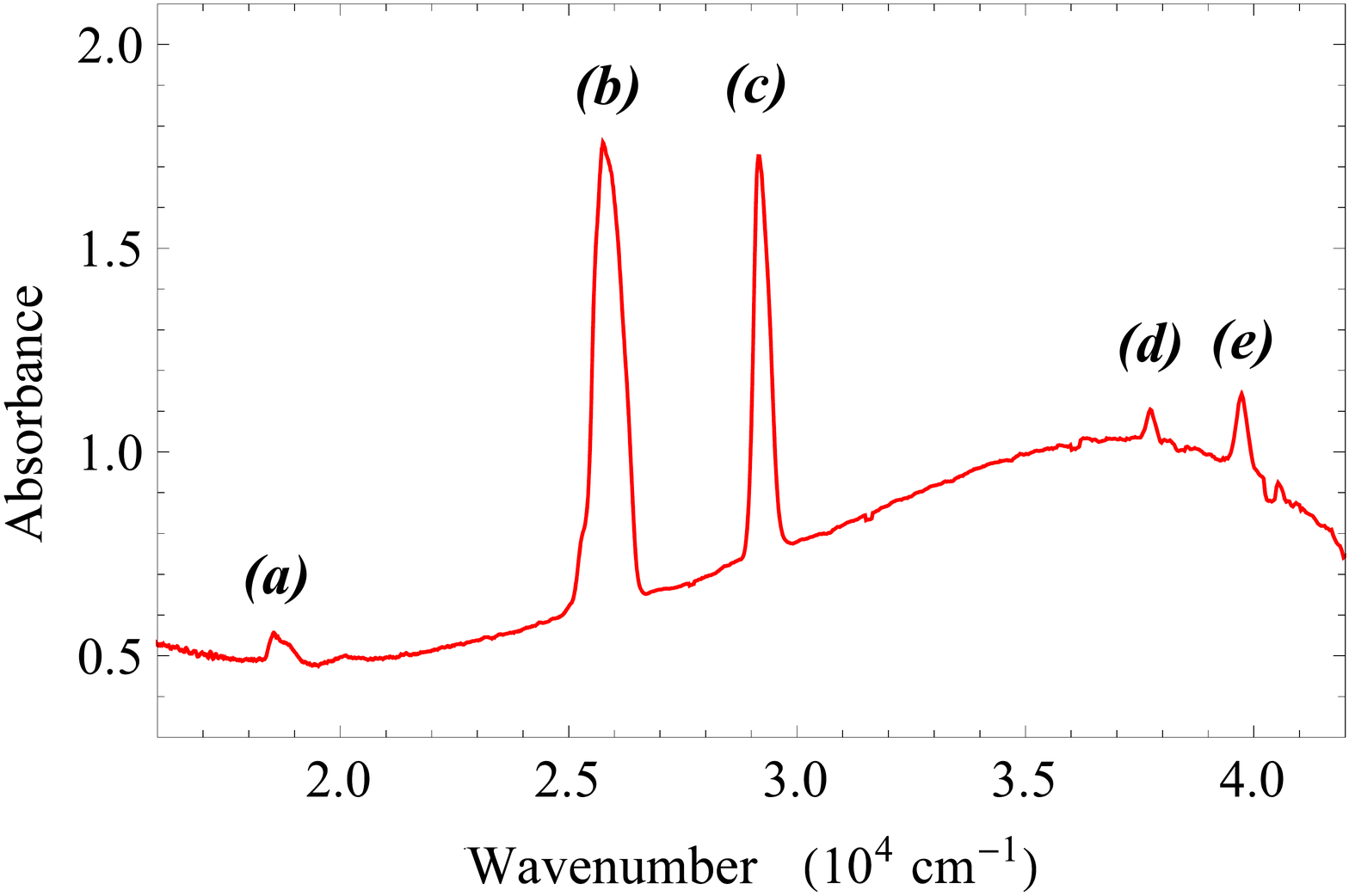}
\caption{The absorption spectrum of Yb atoms in solid Ne taken by the Ocean Optics spectrometer with 1.5 nm resolution.
All five absorption peaks are identified as Yb atomic transitions from the ground state.
See Table \ref{tab:abs} for details.
\label{fig:abs_coarse}}
\end{figure}
\begin{figure*}[t]
\centering
\includegraphics[width=0.9\textwidth]{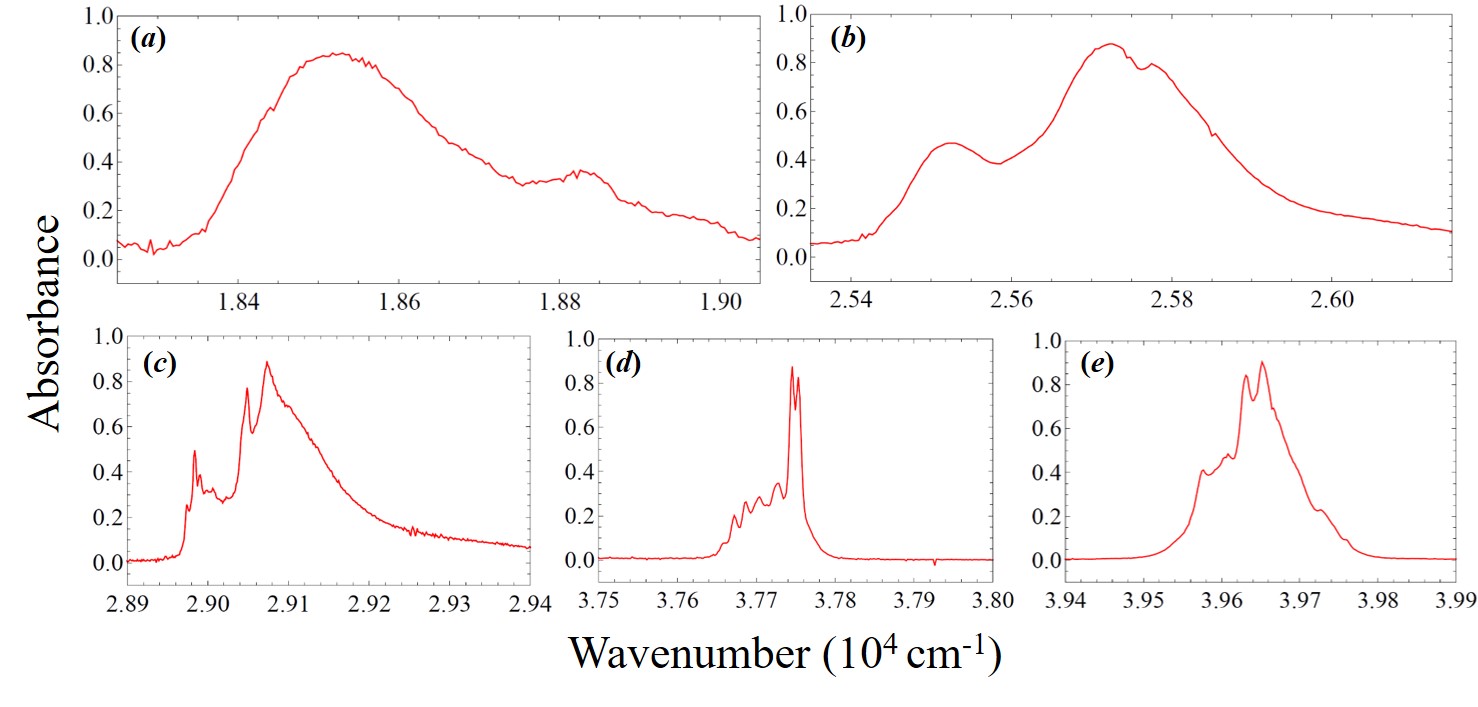}
\caption{The detailed lineshapes of the absorption bands shown in Fig. \ref{fig:abs_coarse}, registered using the McPherson spectrometer at 0.025 nm resolution.
The labels (a)-(e) are the same as those used in Table \ref{tab:abs}.
\label{fig:abs_fine}}
\end{figure*}

Figures \ref{fig:abs_coarse} and \ref{fig:abs_fine} show the white-light absorption spectra of a typical sample taken, respectively, with the 1.5 nm resolution Ocean Optics and 0.025 nm McPherson  spectrometers. Sample deposition leads to a decrease in the transparency of the subtrate which manifests itself as frequency-dependent distortions in the baselines. Nonetheless, five absorption bands, represented in terms of absorbance, can be clearly seen and are identified as due to electronic transitions originating from ground state Yb atoms.
Given that the configuration of the ground state is $4f^{14}5d^06s^2$, low-lying excitations can either be $6s \rightarrow 6p$ outer-shell transitions or $4f \rightarrow 5d$ inner-shell transitions, the latter of which break the full $4f$ shell and leave the $6s$ shell intact.
In the spectra, bands (a) and (b) correspond to outer-shell transitions whose final levels are $6s6p\ ^3\!P_1$ and $6s6p\ ^1\!P_1$ respectively.
In contrast, bands (c), (d) and (e) correspond to inner-shell transitions and their final states are $[4f^{13}]_{j_1}[5d^1]_{j_2}6s^2$ in the $jj$ coupling scheme with $(j_1,j_2)_J$ equal to $(7/2,5/2)_1$, $(5/2,5/2)_1$ and $(5/2,3/2)_1$, respectively.
A comparison between the wavenumbers of transitions in vacuum and those in sNe is summarized in Table \ref{tab:abs}.

\begin{table}[b]
\caption{The assignment of the absorption bands appearing in Fig. \ref{fig:abs_coarse}.
Here, $\sigma_0$ represents the wavenumbers of the transitions in vacuum, $\sigma_\mathrm{abs}$ the wavenumbers of the transitions in solid Ne, and $\Delta\sigma$ the wavenumber shifts. The wavenumbers are in cm$^{-1}$ with an uncertainty of $\pm40$ cm$^{-1}$ in solid Ne.}
\begin{tabular}{c  c  c  c  c}
\hline\hline
Band & Transition & $\sigma_0$ & $\sigma_\mathrm{abs}$ & $\Delta\sigma$\\
\hline
(a) & $6s^2\ ^1\!S_0 \rightarrow 6s6p\ ^3\!P_1$ & 17,992 & 18,510 & +518\\
(b) & $6s^2\ ^1\!S_0 \rightarrow 6s6p\ ^1\!P_1$ & 25,068 & 25,760 & +692\\
(c) & $6s^2\ ^1\!S_0 \rightarrow 4f^{13}5d^16s^2\ (7/2,5/2)_1$ & 28,857 & 29,140 & +283\\
(d) & $6s^2\ ^1\!S_0 \rightarrow 4f^{13}5d^16s^2\ (5/2,5/2)_1$ & 37,415 & 37,730 & +315\\
(e) & $6s^2\ ^1\!S_0 \rightarrow 4f^{13}5d^16s^2\ (5/2,3/2)_1$ & 38,422 & 39,680 & +1,260\\
\hline\hline
\end{tabular}\label{tab:abs}
\end{table}

Band (a), assigned to the intercombination $6s^2\ ^1\!S_0 \rightarrow 6s6p\ ^3\!P_1$ transition, was not observed in our earlier work but becomes visible here due to an increase in the Yb areal density by an order of magnitude \cite{xu11}. Another difference is in our reassignment of band (d). Recent spectroscopic data and calculations of free Yb atoms \cite{wc79,zdwh02,ko10} indicate that it should be associated the $4f^{13}5d^16s^2\ (5/2,5/2)_1$ state rather than, as previously thought, the $4f^{13}5d^26s^1$ one. In the case of band (e), there was also some ambiguity in its assignment originally. In earlier work it has been assigned to to the excited $6s7p\ ^1\!P_1$ level, the energy of which is 40,564 cm$^{-1}$ above the ground state in vacuum  \cite{sa88}. This implies a red-shift in its transition frequency in sNe that is peculiar given that the other four transitions are all blue-shifted and makes the narrowness of its peaks inconsistent with the typical effects of the matrix on an $s \rightarrow p$ transition.
Reassigning it to the $4f^{13}5d^16s^2\ (5/2,3/2)_1$ level removes these inconsistencies, though at the cost of a larger-than-expected absolute shift.

In matrix isolation, outer-shell and inner-shell transitions have different lineshapes due to the screening of medium effects by the outer-shell electrons \cite{sa88}.
For Yb in sNe, the difference can be readily seen by comparing bands (a) and (b) ($6s \rightarrow 6p$) with bands (c), (d) and (e) ($4f \rightarrow 5d$) at the higher spectral resolution of Figure \ref{fig:abs_fine}. Band (a) has two principal structures with maxima at 18,530 cm$^{-1}$ and 18,826 cm$^{-1}$, while band (b) has three principal structures with maxima at 25,525 cm$^{-1}$, 25,724 cm$^{-1}$ and 25,774 cm$^{-1}$. Although both bands (a) and (b) display some splitting, the linewidth of each peak remains a few hundred wavenumbers. 
By contrast, bands (c), (d) and (e) have more complex structures in which the narrowest linewidth is comparable to the instrument resolution. The three principal structures in band (c) have their peaks at 28,988 cm$^{-1}$, 29,049 cm$^{-1}$ and 29,073 cm$^{-1}$. In band (e) there are two clear peaks at 39,631 cm$^{-1}$ and 39,652 cm$^{-1}$ and an unresolved shoulder extending from 39,580 to 39,610 cm$^{-1}$. In band (d) we have a distinct doublet with peaks at 37,741 cm$^{-1}$ and 37,751 cm$^{-1}$ and a series of smaller peaks constituting a substructure that extends to lower frequencies.  

In the subsequent sections, we propose that the principal peaks of these latter bands are connected to ZPLs on individual crystal field levels. We also propose that substructures on these bands are the manifestations of phonon excitations that produce the phonon sidebands typically associated with ZPLs, of which band (d) offers a prime example.

\subsection{Probe with Laser Light}
\label{Sec:Probe_laser}

\begin{figure}[t]
\centering
\includegraphics[width=0.47\textwidth]{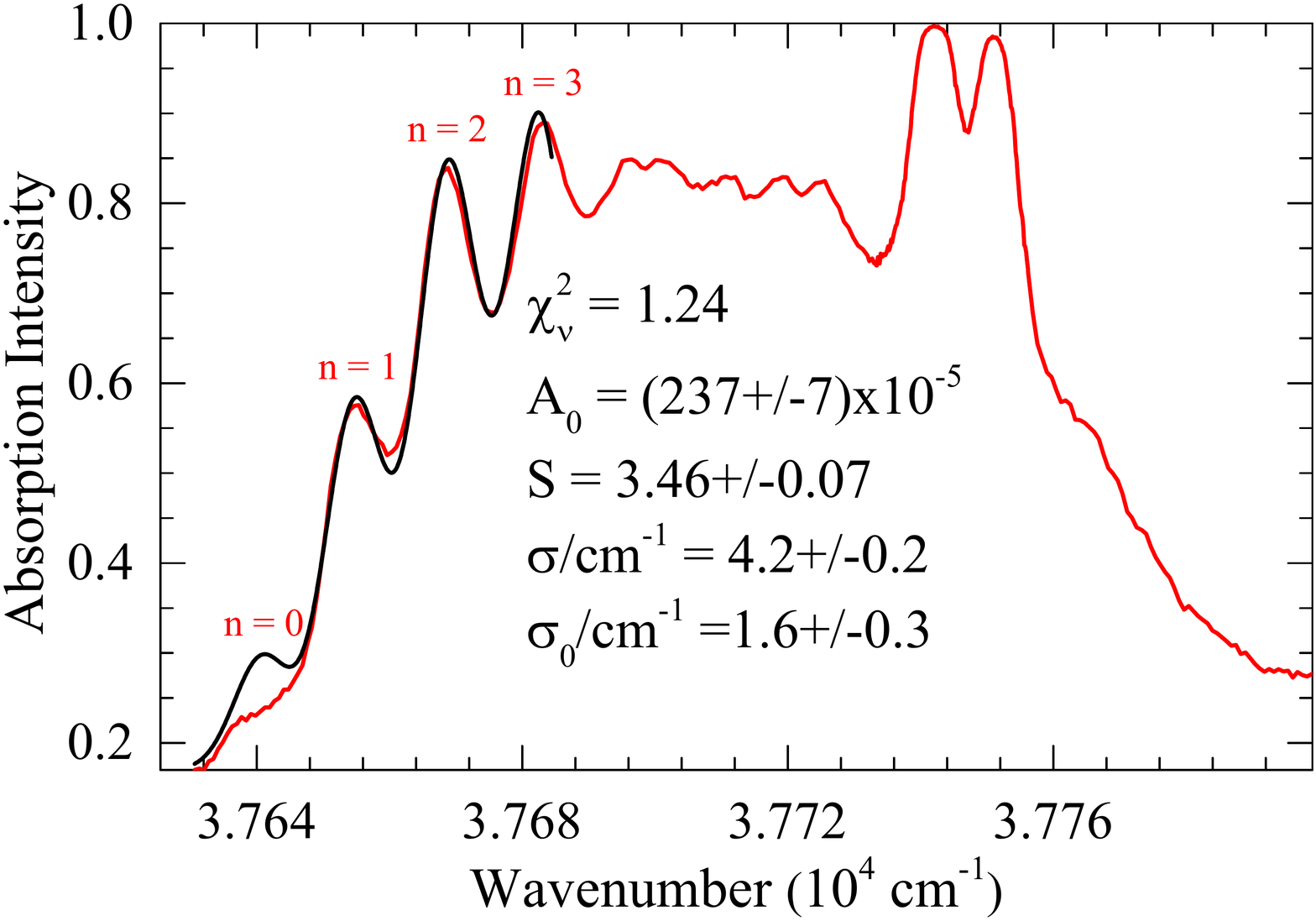}
\caption{Absorption intensity of the $6s^2\ ^1\!S_0 \rightarrow 4f^{13}5d^16s^2\ (5/2,5/2)_1$  transition using the frequency doubled OPO as the probe light.
The spectrum has a resolution of 2 GHz. The parameters obtained by using Equation \eqref{absorption_shape_of_you} to fit the first four features of the vibronic band---indicated by the black line---are given inset.
\label{fig:abs_265}}
\end{figure}

In order to investigate further the vibronic progression seen in band (d), the $6s^2\ ^1\!S_0 \rightarrow 4f^{13}5d^16s^2\ (5/2,5/2)_1$ transition was probed at 2 GHz resolution using frequency doubled light from the OPO. Figure \ref{fig:abs_265} shows a plot of the absorbed light intensity (normalized against the incident light intensity) versus wavenumber. As with the original white-light spectra, the two peaks centered at 37,741 cm$^{-1}$ and 37,751 cm$^{-1}$ were assigned to the $4f^{13}5d^16s^2\ (5/2,5/2)_1 $ state. Extending to lower frequencies, between 37,631 cm$^{-1}$ and 37,733 cm$^{-1}$, one can see a series of 9 smaller, consecutive peaks.  Possible explanations for this series include multiple site occupancy and/or CFS. However, both of these seem unlikely in this case. First, there is no analogous structure in bands (c) and (e), which indicates that this is a phenomenon peculiar to the $4f^{13}5d^16s^2\ (5/2,5/2)_1$ transition. Second, attributing this series to multiple site occupancy would require the existence of too many additional trapping sites: at least 3 of axial symmetry, given that CFS gives rise to 3 features for each $J=1$ state; and at most 9, if we assume a one-to-one correspondence between the peaks and trapping sites. This is unsupported by the theoretical calculations of Section \ref{Sec:Sites}, while the regularity of the peaks---a spacing of 14 cm$^{-1}$ for the first four---has never been observed in other cases of multiple site occupation. 
  
Instead, we interpret the harmonic series of Fig. \ref{fig:abs_265} as evidence of coupling between the electronic transition and a small number of local phonon modes. This type of structure is usually not visible in RG matrices as electronic transitions tend to couple strongly to a large number of lattice modes to produce broad unstructured bands in the spectra. In ionic crystals, the emergence of strong vibronic sidebands in intra 4f$^N$ electronic transitions is associated with strong configuration interaction and coupling to charge transfer states \cite{gu14}. These mechanisms likely also operate here since crystal field splitting induced by van der Waals forces in the neutral Mn/Kr system is attributed to the dominance of: (i) inter-electronic repulsion in the metal atom in the case of weak splitting; or (ii) ligand-metal electron repulsion in the case of strong splitting \cite{by10}.   

In the most common treatment, the strength of the electron-phonon coupling is measured by the Huang-Rhys parameter, $S$, which is related to the displacement in the lattice equilibrium position near a defect site, $\Delta$, accompanying an electron transition according to: 
\begin{equation}
\label{yellow}
S = \frac{4\beta^2}{(1+\beta^2)^2}\frac{\omega M \Delta^2}{2\hbar}, \ \beta^2 = \frac{\omega^{\prime}}{\omega},
\end{equation}
for which $\omega$ and $\omega^{\prime}$ are the frequencies of the vibrational mode in the initial and final electronic states, respectively, and $M$ is the effective mass of the defect site. At very low temperature (i.e. $\hbar\omega > kT \sim 2.8 - 3.5$ cm$^{-1}$) vibronic transitions can be approximated as arising from the zero-phonon ($n =0 $) state and $p_n$, the probability that the $n$th vibrational state is populated, can be straightforwardly shown under the Franck-Condon approximation to obey $p_n(S) \propto \exp(-S)S^n/n!$ which is also the form of the overall envelope of the vibronic band \cite{zh19}. Within that band, on the assumption that coupling involves a single vibrational mode and that the frequencies of the oscillators follow a Gaussian distribution around the mode's average frequency, $\omega$, the intensity function of the phonon line $n$ is \cite{wa64}: 
\begin{equation}
\label{shape_of_you}
\begin{split}
f_n(E) & = \exp(-S)\frac{S^n}{n!} \\
& \times \frac{1}{\sqrt{2\pi n\sigma^2}}\exp \left(\frac{-[E-(E_0+n\hbar\omega)]^2}{2n\sigma^2} \right),
\end{split}
\end{equation}
for which $\sigma\sqrt{n}$ is its linewidth and $E_0$ is the energy difference between the initial and final electronic states. 

We therefore expect to see a harmonic progression in the absorption spectrum in intervals of $\hbar\omega$ and, in Fig. \ref{fig:abs_265}, the three phonon lines corresponding to $n = 1, 2, 3$ can be clearly distinguished, centered at $37,655$ cm$^{-1}$, $37,669$ cm$^{-1}$ and $37,683$ cm$^{-1}$, respectively. Per Equation \eqref{shape_of_you}, the $n =0$ order corresponding to the ZPL should be a $\delta$-function. In reality, because of electronic dispersion and inhomogeneous broadening due to trapping site inhomogeneity it has a finite width, $\sigma_0$. Higher order ($n > 3$) phonon lines are also visible in the range of 37,690 cm$^{-1}$ to 37,733 cm$^{-1}$, though they are not so easily resolved, possibly because the single phonon approximation no longer holds and interference between multiple modes should be considered. For the sake of simplicity, we only fit the $n = 0-3$ orders using a function for the absorption intensity, $A(E)$, that superimposes the ZPL on the single frequency mode harmonic progressing band \cite{gu13}:   
\begin{equation}
\label{absorption_shape_of_you}
\begin{split}
A(E) &= A_0\sum_{n=0}^{n=3}\exp(-S)\frac{S^n}{n!}\left(\frac{E_0+n\hbar\omega}{E_0} \right) \\
&\times \frac{1}{\sqrt{2\pi(\sigma_0^2+n\sigma^2)}}\exp\left(-\frac{[E-(E_0+n\hbar\omega)]^2}{2(\sigma_0^2+n\sigma^2)} \right) + O.
\end{split}
\end{equation} 

The small fixed offset $O = A(E= 37,630$ cm$^{-1}) = 0.1703$ accounts for the non-zero background in Fig. \ref{fig:abs_265} produced by light scattered from the matrix. The same figure gives the phonon and ZPL frequencies as $\omega = 14.0$ cm$^{-1}$ and $E_0 =  37,641$ cm$^{-1}$, respectively. The black trace shows that the fit was performed in the range of 37,630 to 37,685 cm$^{-1}$ and gives the following values for the parameters: $A_0 = (237\pm 7)\times10^{-5}$, $S = 3.46\pm 0.07$,  $\sigma_0 = 1.6\pm 0.3$ cm$^{-1}$ and $\sigma = 4.2\pm 0.2$ cm$^{-1}$. The lattice displacement, $\Delta$, can be calculated using Equation \eqref{yellow} under the approximations that $\beta \simeq 1$ (i.e. $\omega \simeq \omega^{\prime}$) and $M$ is equal to the sum of the masses of the Yb atom and its nearest neighbor Ne atoms. We thus obtain $\Delta = 1.29\pm 0.01$  {\AA}, which is approximately 25\% of the ground-state Yb--Ne bond length. This displacement is quite realistic and supports our value for $S$. It can be easily seen that the fit is least accurate for the $n=0$ feature. The ZPL  at $E_0 =  37,641$ cm$^{-1}$ is only minimally visible because its intensity is reduced in favor of the phonon sideband. It should be more prominent at lower temperatures. The maxima of the peaks at 37,741 cm$^{-1}$ and 37,751 cm$^{-1}$ cannot be completely discerned due to saturation of the detector. However, their sharpness in Fig. \ref{fig:abs_coarse} (d) strongly suggests that they are also ZPLs. 

\subsection{Emission Spectroscopy}

%
In earlier work, we showed that strong fluorescence of Yb atoms trapped in sNe can be induced by driving the $6s^2\ ^1\!S_0 \rightarrow 6s6p\ ^1\!P_1$ transition \cite{Lambo_2012,xu14}.
Both a 1 MHz linewidth laser centered at 388 nm (25,770 cm$^{-1}$) and a 10 nm linewidth LED centered at 385 nm (25,970 cm$^{-1}$) were able to resonantly excite this transition.
This results in a significant population transfer to the metastable  $6s6p\ ^3\!P_0$ level via $6s6p\ ^1\!P_1 \rightarrow 6s5d\ ^3\!D_J$ intersystem crossing, the branching ratio of which is enhanced by level mixing induced by the crystal field of the matrix.
Due to the long lifetime of the $6s6p\ ^3\!P_0$ state in sNe, excited Yb atoms continue absorbing photons to make transitions from this state to even higher excited ones.
A strong absorption peak centered at 26,740 cm$^{-1}$, corresponding to the $6s6p\ ^3\!P_0 \rightarrow 6p^2\ ^3\!P_1$ excitation, has been previously reported \cite{xu11}. Given the broad linewidths of absorption peaks in the solid state, the 385 nm LED can also excite this transition and emission from states at higher energies than the exciting photons are thus observed.
%
%
\begin{table}[b]
\caption{The assignment of identifiable emission lines appearing in Fig. \ref{fig:emis_fine}.
Transitions forbidden in vacuum are indicated by the symbol $\dagger$.
Five of the emissions lines are correlated to absorption lines listed in Table \ref{tab:abs} and are labeled with the same letters.
Here, $\sigma_0$ represents the wavenumbers of transitions in vacuum, $\sigma_\mathrm{emis}$ the wavenumbers of the transitions in solid Ne, and $\Delta\sigma$ the wavenumber shifts.
The wavenumbers are in cm$^{-1}$ with an uncertainty of $\pm40$ cm$^{-1}$ in solid Ne.}
\begin{tabular}{c c c c c}
\hline\hline
Transition & $\quad$ & $\sigma_0$ & $\sigma_\mathrm{emis}$ & $\Delta\sigma$\\
\hline
$6s6p\ ^3\!P_0 \rightarrow 6s^2\ ^1\!S_0$ & $(i)^{\dagger}$  & 17,288 & 17,730 & $+440$\\
$6s6p\ ^3\!P_1 \rightarrow 6s^2\ ^1\!S_0$ & $(ii)/(a)$ & 17,992 & 18,320 & $+320$\\
$6p^2\ ^3\!P_1 \rightarrow 6s6p\ ^1\!P_1 $ & $(iii)$ & 18,737 & 18,900 & $+160$\\
$6s6p\ ^3\!P_2 \rightarrow 6s^2\ ^1\!S_0$ & $(iv)^{\dagger}$ & 19,710 & 20,080 & $+370$\\
\hline
$4f^{13}5d^16s^2\ (7/2,3/2)_2 \rightarrow 6s^2\ ^1\!S_0$ & $(v)^{\dagger}$ & 23,189 & 23,070 & $-120$\\
$6s5d\ ^3\!D_1 \rightarrow 6s^2\ ^1\!S_0$ &  $(vi)^{\dagger}$ & 24,489 & 24,490 & $\sim0$\\
$6s6p\ ^1\!P_1 \rightarrow 6s^2\ ^1\!S_0 $ & $(vii)/(b)$ & 25,068 & 25,320 & $+250$\\
\hline
$6s5d\ ^1\!D_2 \rightarrow 6s^2\ ^1\!S_0 $ & $(viii)^{\dagger}$ & 27,678 & 27,860 & $+180$\\
$4f^{13}5d^16s^2\ (7/2,5/2)_1 \rightarrow 6s^2\ ^1\!S_0 $ & $(ix)/(c)$ & 28,857 & 29,030 & $+170$\\
$4f^{13}6s^26p\ (7/2,3/2)_2 \rightarrow 6s^2\ ^1\!S_0 $ & $(x)^{\dagger}$ & 35,197 & 35,590 & $+390$\\
$4f^{13}5d^16s^2\ (5/2,5/2)_1 \rightarrow 6s^2\ ^1\!S_0 $ & $(xi)/(d)$ & 37,415 & 37,730 & $+320$\\
$4f^{13}5d^16s^2\ (5/2,3/2)_1 \rightarrow 6s^2\ ^1\!S_0 $ & $(xii)/(e)$ & 38,422 & 39,610 & $+1,190$\\
\hline\hline
\end{tabular}\label{tab:emis}
\end{table}

\begin{figure*}[t]
\centering
\includegraphics[width=0.9\textwidth]{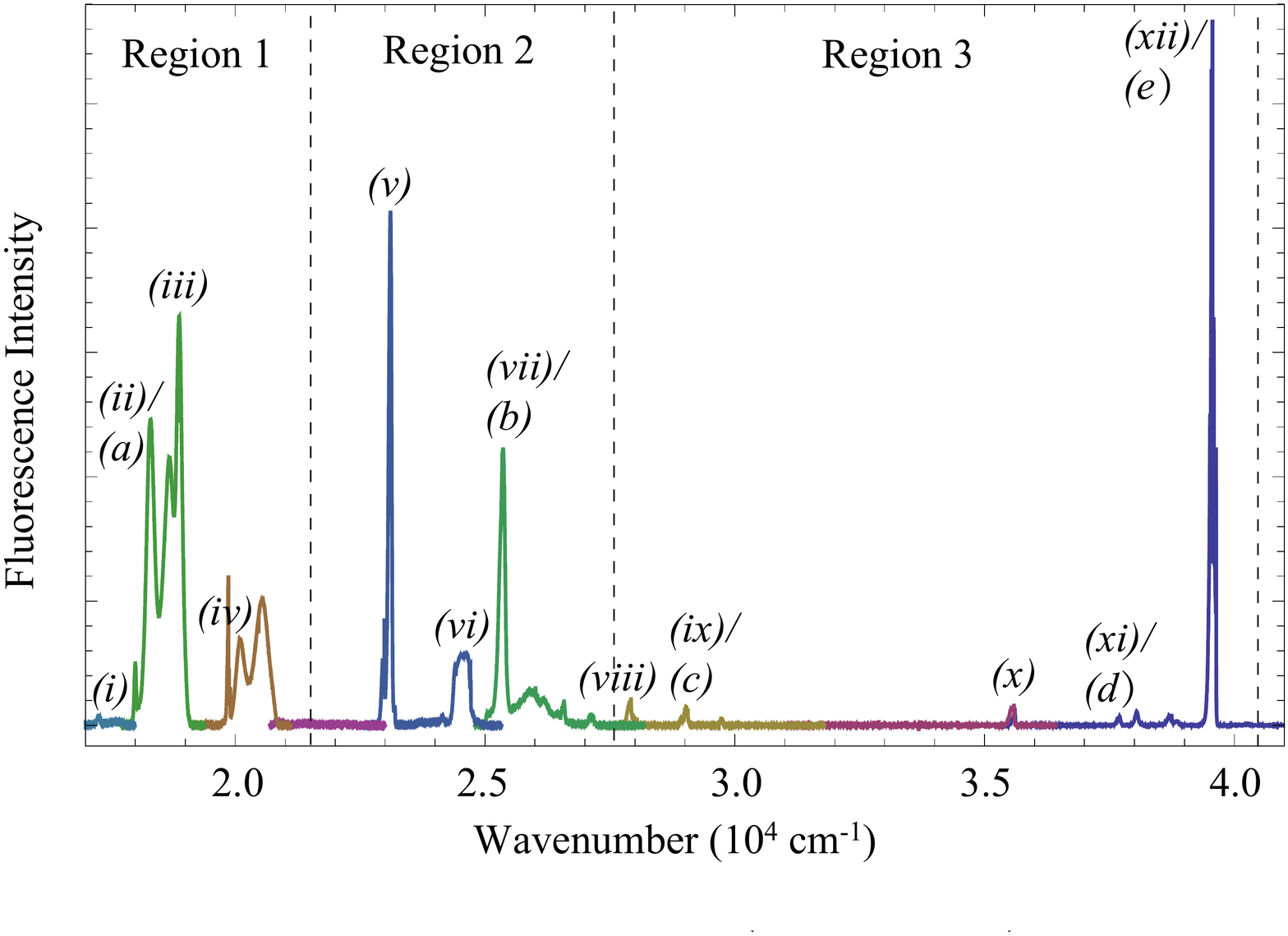}
\caption{The steady-state emission spectrum of Yb atoms in solid Ne induced by the 385 nm LED and recorded using the McPherson spectrometer at 0.025 nm resolution. The extent of each color represents the extent of the camera frame. The labels on the identified emission lines are the same as those used in Table \ref{tab:emis}. 
\label{fig:emis_fine}}
\end{figure*}

Figure \ref{fig:emis_fine} shows the full steady state emission spectrum of Yb atoms in sNe excited by the 385 nm LED and recorded at a resolution of 0.025 nm. We divide the emission lines into the three wavenumber regions given below and summarize those we were able to identify in Table \ref{tab:emis}.
%

{\bf Region 1} ($1.7\sim2.1\times10^{4}$ cm$^{-1}$): The prominent emission lines in this region are responsible for a green glow from the sample when its is pumped by the LED.
They include the decays of the fine-structure triplet group $6s6p\ ^3\!P_{0,1,2} \rightarrow 6s^2\ ^1\!S_0$ whose spacing in sNe is approximately the same as in the free atom.
Although metastable in vacuum, the $6s6p\ ^3\!P_{0,2}$ levels decay radiatively in matrix isolation due to a Stark-like mixing induced by the crystal field. The assignments of the $6s6p\ ^3\!P_{0,1}$ states have been validated by lifetime measurements in earlier work \cite{xu14}. It should be noted that the strong emission line at 18,900 cm$^{-1}$ does not belong to this triplet group but corresponds to the $6p^2\ ^3\!P_1 \rightarrow 6s6p\ ^1\!P_1$ decay following the $6s6p\ ^3\!P_0 \rightarrow 6p^2\ ^3\!P_1$ excitation mentioned above \cite{xu11}. The emission line at 20,710 cm$^{-1}$ is the most prominent one that remains unidentified in this region.

{\bf Region 2} ($2.2\sim2.6\times10^{4}$ cm$^{-1}$):
Aside from the $6s6p\ ^1\!P_1$ decay at 25,320 cm$^{-1}$, two strong emission lines are seen in this region. We tentatively propose that the line at 23,070 cm$^{-1}$ corresponds to the decay of the $4f^{13}5d^16s^2\ (7/2,3/2)_2$ state. This excited level has the same electronic configuration as those of absorption peaks (c), (d) and (e), but the $jj$ coupling of $j_1=7/2$ and $j_2=3/2$ does not give $J=1$ and its decay is thus forbidden in vacuum. This assignment results in the only negative matrix shift in Table \ref{tab:emis}. However, attempts to assign it to other nearby states, such as $6s5d\ ^3\!D_{1,2}$, result in unnaturally large matrix shifts. On similar grounds, the parity forbidden $6s5d\ ^3\!D_1 \rightarrow 6s^2\ ^1\!S_0$ transition is proposed to be responsible for the 24,490 cm$^{-1}$ line. 

{\bf Region 3} ($2.7\sim4.0\times10^{4}$ cm$^{-1}$):
Emission lines in this region have photon energies higher than the exciting photon. We again assign the decay of various excited states according to the proximity of their wavenumbers to those of excited states obtained in vacuum. Thus, the 28,857 cm$^{-1}$ emission corresponds to $4f^{13}5d^16s^2\ (7/2,5/2)_1$ decay; the 37,730 cm$^{-1}$ emission corresponds to $4f^{13}5d^16s^2\ (7/2,5/2)_1$ decay; and the 39,610 cm$^{-1}$ emission corresponds to $4f^{13}5d^16s^2\ (7/2,5/2)_1$ decay. The profile of the absorption energies that populate these excited states are, respectively, peaks (c), (d) and (e) of Fig. \ref{fig:abs_fine}. Although the matrix shift of the $6s^2\ ^1\!S_0 \rightarrow 4f^{13}5d^16s^2\ (5/2,3/2)_1$ transition appears rather large, it is comparable to the size of the matrix shift that occurs in the absorption spectrum.

%

\section{Discussion}
\label{Sec:Discussion}

In order to interpret the structure of the absorption bands, we discuss here the consistency of all experimental data invoking also the results of limited theoretical simulations possible for the Yb/Ne system.

\subsection{Stable trapping sites}
\label{Sec:Sites}

The thermodynamically stable trapping sites of ground-state atomic Yb in a perfect face-centered cubic (fcc) Ne crystals were  modeled following the approach suggested in our earlier works~\cite{Tao2015,kleshhina19Yb,kleshchina19Ba}. The full details of the simulations are summarized in the Supplemental Materials to this paper. In brief, we used {\it ab initio} Yb($^1S_0$)--Ne \cite{Lambo_2012} and slightly modified Aziz-Slaman Ne--Ne potentials \cite{Aziz1989} to represent the pairwise force field. For a large fragment of the Ne crystal, composed of more than 3,000 atoms, the Yb accommodation energy, $\Delta E(N)$, was obtained as a function of $N = 0, 1,... 20$, the number of Ne atoms removed from the crystal. For each $N$, the lowest energy was found by minimization of the trapping site structure in configuration space and corrected by the energy required to remove $N$ crystal atoms. 

\begin{figure}[t]
\begin{center}
	\includegraphics[width=0.48\textwidth]{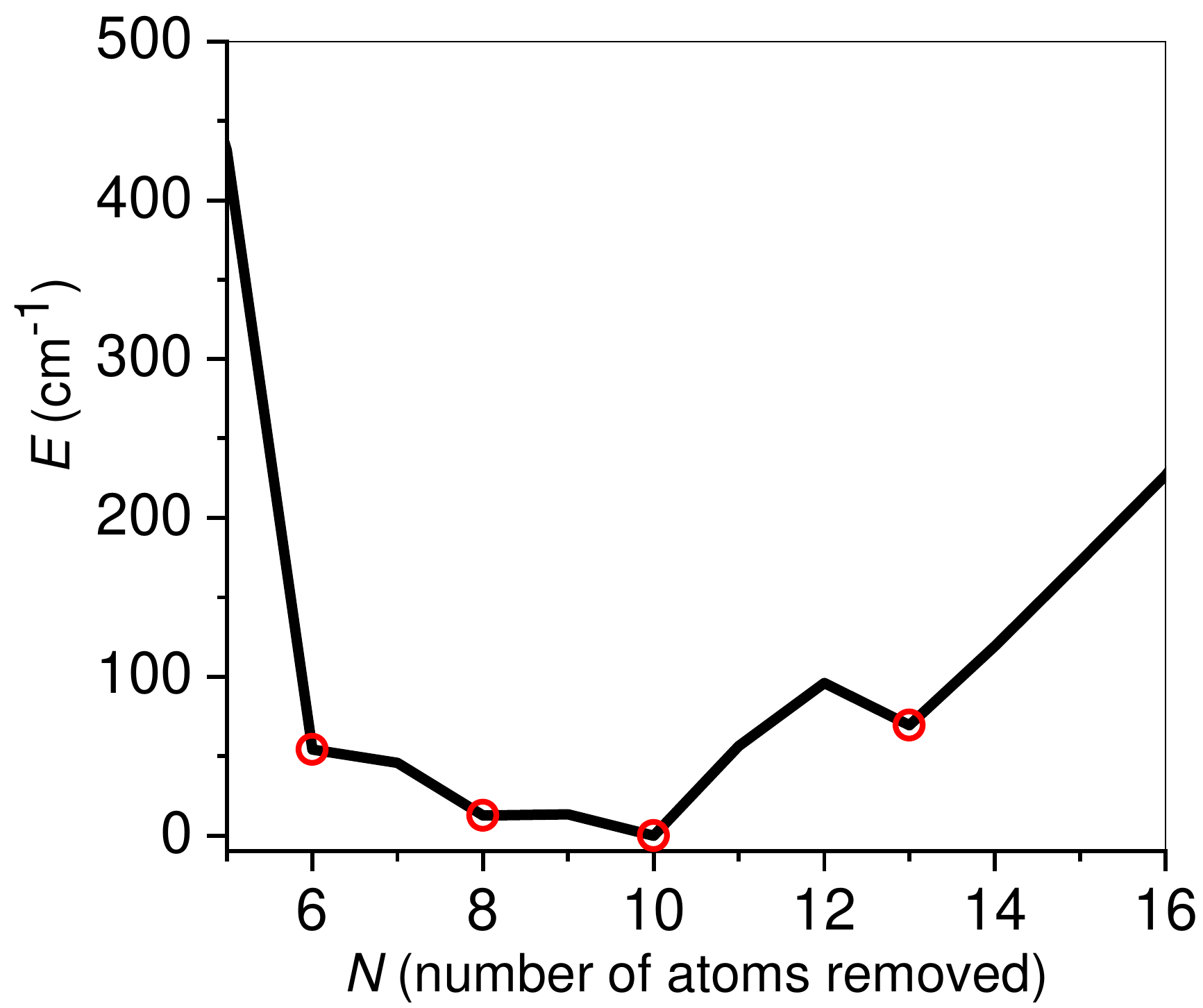}
\includegraphics[width=0.4\textwidth]{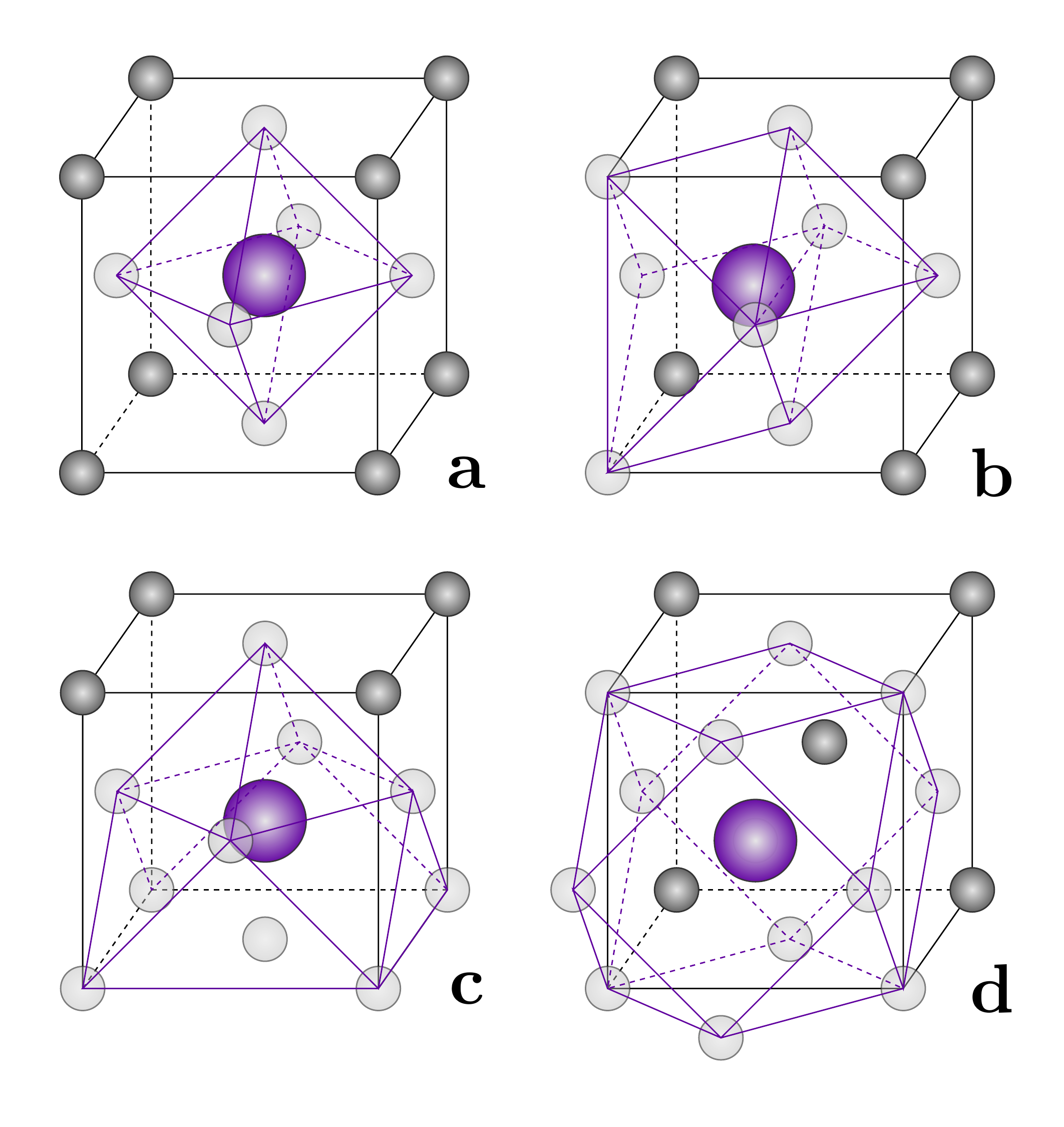}
\caption{(color online) Above, the accommodation energy of the Yb atom in the Ne fcc crystal as a function of $N$, the number of Ne atoms removed from the lattice (solid line). Red circles indicate the points lying on the convex hull of the $E(N)$ diagram. Below, the schematic structures of the Yb/Ne stable trapping sites, 6V (a), 8V (b), 10V (c) and 13V (d). Yb and Ne atoms are represented by large and small spheres, respectively; light grey spheres indicate the lattice positions of Ne atoms removed from the system; dark grey spheres indicate the lattice positions of the remaining Ne atoms.
}\label{fig:convex}
\end{center}
\end{figure}

The meaningful part of the resulting $E(N)$ diagram is shown in Fig.~\ref{fig:convex}. 
It identifies thermodynamically stable sites as those whose energies lie on its convex hull, in close analogy to the analysis of the discrete variable composition phase diagrams~\cite{Zhu2013}. Four such sites are evident: the ground one, having the lowest accommodation energy, with Yb in an $N=10$ vacancy and three others lying higher in energy with Yb in $N=8,\ 6$ and 13 vacancies (referred to in what follows as $N$V for ``$N$-atom vacancy''). Thus, the Ne crystal tends to accommodate the Yb atom in quite spacious multiple-atom vacancies. By contrast, $N=4$ and $N=6$ sites compete with each other in solid Ar, an $N=4$ site dominates in solid Kr, and a single substitutional $N=1$ site is the most stable one in solid Xe \cite{kleshhina19Yb}. Moreover, the energies of the perfect octahedral site, 6V, and the cuboctahedral site, 13V, are quite large, while the more energetically stable 10V and 8V sites have only axial coordination symmetries, $C_{4v}$ and $C_{2v}$, respectively (see Supplemental Materials). 

These findings have important implications for the CFS of the absorption bands related to $J=1$ level excitations. In the rigid crystal environment of $C_{4v}$ symmetry, they should appear as two degenerate ZPLs and a single ZPL (with associated substructures if multiple phonons are excited) according to the projection of $J$ onto the site axis $\Omega=\pm 1$ and 0, falling into $E$ and $A_1$ irreducible representations, respectively. The electron-phonon Jahn-Teller interaction should lift the remaining degeneracy, producing a 2+1 or 1+2 generic band structure, as was observed for the axial trapping site of the Ba atom \cite{dm18,dgm18,kleshhina19YbBa}. For the site of $C_{2v}$ symmetry, CFS produces three individual ZPLs.   

\subsection{$6s \rightarrow 6p$ Absorption Spectra}
\label{Sec:spabsorption} 

The absorption bands due to $6s \rightarrow 6p$ transitions in Figs. \ref{fig:abs_fine} (a) and (b), appear as broad, weakly structured features, typical of allowed or weakly forbidden outer-electron transitions ~\cite{by10,dm16,dm18}. Upon $s \rightarrow p$ electron promotion, the atom-matrix interaction changes dramatically and becomes strongly anisotropic, so that multiple phonon excitations superimpose relatively large crystal field splittings. At a glance, the $6s^2\ ^1\!S_0 \rightarrow 6s6p\ ^3\!P_{1}$ and $6s^2\ ^1\!S_0 \rightarrow 6s6p\ ^1\!P_{1}$ transition profiles of Figs. \ref{fig:abs_fine} (a) and  \ref{fig:abs_fine} (b), respectively, can be interpreted as 2+1 and 1+2 split bands originating from the ground 10V site.  

\begin{figure}[h]
\begin{center}
	\includegraphics[scale=0.38]{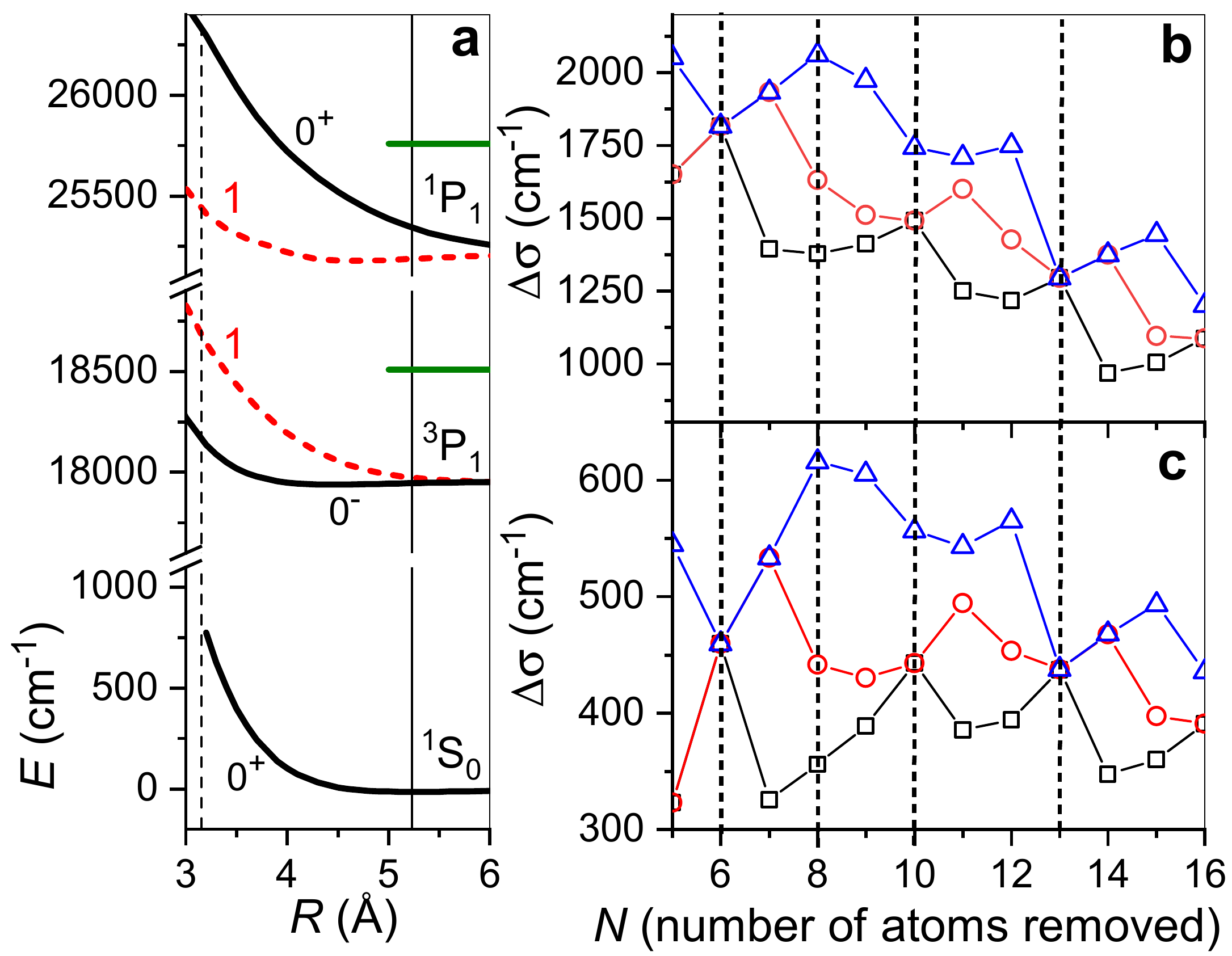}
\caption{(color online) (a) Yb-Ne interaction potentials for the $^3\!P_1$ and $^1\!P_1$ atomic states as functions of the internuclear distance $R$ with $\Omega$ values indicated. The solid vertical line corresponds to the ground-state equilibrium distance, and the dashed vertical line to the Ne-Ne distance in the fcc crystal. The green horizontal bars on the right indicate the maxima of the corresponding absorption bands in Figs. \ref{fig:abs_fine}. (b) Vertical $6s^2 \, {^1}S_0 \to 6s6p \, ^1P_1$ transition frequency shifts for the lowest energy structures with different $N$. The shifts for each of the excited-state adiabatic PESs relative to the transition frequency in vacuum are presented in black, red and blue colors in ascending order. The vertical dashed lines indicate thermodynamically stable sites. (c) For the $6s^2 \, {^1}S_0 \to 6s6p \, ^3P_1$ transition the same is shown as for the $6s^2 \, {^1}S_0 \to 6s6p \, ^1P_1$ transition in (b). 
}\label{fig:vertical}
\end{center}
\end{figure}

To verify this interpretation, we simulated the $6s \rightarrow 6p$ absorption spectra using the site geometries found for the ground state and the diatomics-in-molecule model \cite{Ryan2010,Boatz1994,Visticot1994,Kryloval1996} parametrized by the Yb($^3\!P_1$,$^1\!P_1$)--Ne {\it ab initio} potentials available in Ref.~\citenum{Lambo_2012}. For the $^3\!P_J$ multiplet, we assumed a weak crystal field limit noting that its spin-orbit coupling constant is on the order of 800 cm$^{-1}$, while the relative matrix-induced shifts in different types of sites do not exceed 100 cm$^{-1}$ (Fig.~\ref{fig:vertical}).  Further details are presented in the Supplemental Materials to this paper. Figure~\ref{fig:vertical}(a) shows the spin-orbit-coupled interaction potentials for the $^3\!P_1$ and $^1\!P_1$ atomic states. The latter state exhibits much stronger interaction anisotropy---i.e. the difference between the potentials for $\Omega=0$ and $\pm$1 components ---than the former state. Moreover, the sign of the anisotropy for each state is the opposite of the other state's. Nevertheless, for both states, strong repulsion of one $\Omega$ component overpowers weak attraction of the other $\Omega$ component and leads to an overall positive matrix shift in their absorption frequencies. 

Panels (b) and (c) of Fig.~\ref{fig:vertical} present the vertical transition frequency shifts relative to the corresponding atomic transition in vacuum for the lowest energy sites at each $N$. Those corresponding to the stable 6V, 8V, 10V and 13V sites are marked by vertical lines. Three adiabatic potentials, which correlate to atomic and diatomic terms $(J=1,\Omega)$, split out differently depending on the crystal field symmetry of the site: the polyhedral 6V and 13V sites maintain threefold degeneracy; the 10V site of $C_{4v}$ symmetry provides a 2+1 degeneracy pattern; and the 7V site of $C_{3v}$ symmetry, which is unstable for Yb in solid Ne but stable for Ba in solid Ar, Kr and Xe \cite{kleshhina19YbBa}, produces 1+2 splitting. In the sites of lower symmetries, degeneracy is completely lifted. Despite anisotropies of opposite signs, the CFS patterns of the singlet and triplet states are qualitatively similar, though the larger splittings in the case of the singlet state reflect its larger interaction anisotropy.

Molecular dynamics simulations were performed for the $6s \rightarrow 6p$ absorption bandshapes, as described in the Supplemental Materials. Figure~\ref{fig:bands} compares the bands simulated for each stable trapping site with bands (a) and (b) of Fig. \ref{fig:abs_fine}. The simulations systematically overestimate the matrix shift of the $6s^2\ ^1\!S_0 \rightarrow 6s6p\ ^1\!P_{1}$ absorption and, to a lesser extent, underestimate the shift of the $6s^2\ ^1\!S_0 \rightarrow 6s6p\ ^3\!P_{1}$ absorption. These errors are attributed to uncertainties in the short-range repulsive branches of the diatomic potentials, which strongly affect the transition frequencies. In agreement with the above analysis of the vertical transition frequencies, the simulated $6s^2\ ^1\!S_0 \rightarrow 6s6p\ ^1\!P_{1}$ bands are very broad and exhibit a remarkable dependence on the symmetry of the trapping site. The structures seen in their shapes fully agree with the CFS of the vertical transitions: 2+1 for the 10V site with the two-fold degeneracy broken by the Jahn-Teller effect; a symmetric Jahn-Teller triplet for the triply degenerate 6V site; and an asymmetric triplet for the 8V site. An unstructured asymmetric lineshape is obtained for the polyhedral 13V site due to the large dynamical distortion of this spacious and labile structure. By contrast, the simulations of the $6s^2\ ^1\!S_0 \rightarrow 6s6p\ ^3\!P_{1}$ absorption predict much narrower and strongly overlapping bands. The smaller anisotropy of this state leads to smaller CFS, which is partially washed out for the 8V site. The Jahn-Teller structure is not resolved at all and only contributes to the broadening of the bands. 

The simulations confirm that the observed $6s^2\ ^1\!S_0 \rightarrow 6s6p\ ^3\!P_{1}$ absorption band bears a 2+1 CFS structure typical of axially-symmetric sites. It may be due to occupation in a single site, most likely the 10V one, or involve contributions from the other trapping sites, which all produce bands of very similar shape, as can be seen in Fig.~\ref{fig:bands}(a). Interpretation of the $^1\!P_1$ absorption is less obvious. Uncertainty in the excited-state Yb($^1\!P_1$)--Ne potentials is unlikely to be so large as to cause an inverted CFS structure for the 10V site, and we therefore explain the observed 1+2 structure as a result of multiple site absorptions, which, for this particular transition, produce the distinct band shapes seen in Fig.~\ref{fig:bands}(b). Attempts to fit the observed envelopes of the absorption spectrum to the weighted sum of the simulated ones gave ambiguous results. Excitation spectroscopy, which we reserve for future work, is expected be more informative for the discrimination of the distinct trapping sites.

\begin{figure}[h]
\begin{center}
	\includegraphics[scale=0.6]{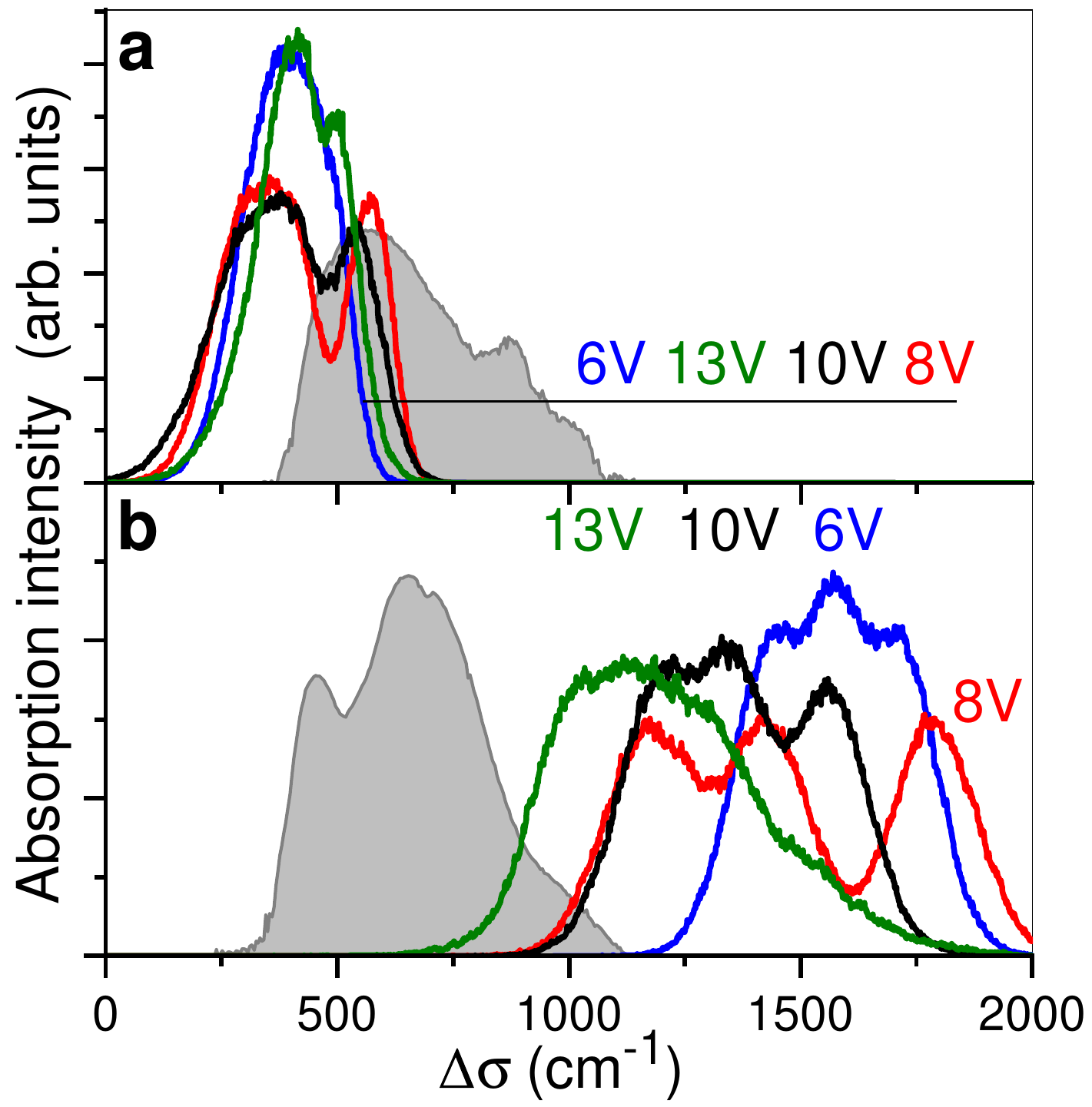}
\caption{(color online) Simulated and observed absorption intensities of the $^3\!P_1$ (a) and $^1\!P_1$ (b) transitions. Frequency is presented in terms of the shift relative to the atomic transition in vacuum. Each band is normalized to unit area. 
}\label{fig:bands}
\end{center}
\end{figure}

\subsection{ $4f \rightarrow 5d$ Absorption Spectra}

The high-resolution absorption spectra of the Yb/Ne system shown in Fig. \ref{fig:abs_fine} exhibit a number of features that have not been previously detected in heavier RG matrices. In the latter, at low resolution, the spectra are structured in a way that reflects occupation in multiple trapping sites. Here, however, we observe much sharper structures that have qualitative similarities across all three $4f \rightarrow 5d$ absorption bands of Figs.~\ref{fig:abs_fine} (c)-(e). As described in Sec. \ref{Sec:Probe_white}: band (c) has a low frequency peak at 28,988 cm$^{-1}$ and a high-frequency doublet with peaks at 29,049 cm$^{-1}$ and 29,073 cm$^{-1}$; band (e) also has a doublet, with peaks at 39,631 cm$^{-1}$ and 39,652 cm$^{-1}$, and a less resolved shoulder to its red from 39,580 to 39,610 cm$^{-1}$; and in band (d) the doublet has its peaks at 37,741 cm$^{-1}$ and 37,751 cm$^{-1}$. The latter band has a long harmonic progression in its red wing, which Fig.~\ref{fig:abs_265} shows to be especially well-resolved under laser excitation. We attributed this progression to phonon sidebands and the fit presented in Sec.~\ref{Sec:Probe_laser} fully confirms this assignment. A more erratic secondary structure seen in band (c) may have the same origin, though the spacings between the first four peaks vary from 6 to 10 cm$^{-1}$. 

It is extremely difficult to address the $4f^{13}5d^16s^2$ states of Yb by means of {\it ab initio} methods and thus perform a theoretical analysis similar to that presented above for the $6s \rightarrow 6p$ absorption spectra. Nonetheless, it is possible to make plausible inferences. First, the structure of the atomic states originating from the $4f^{13}5d^16s^2$ configuration is dominated by $jj$-coupling and shows large spin-orbit splitting~\cite{wc79,zdwh02,ko10}. The weak-field approximation used for the $^3P_J$ multiplet should be equally valid. It is thus reasonable to attribute the overarching band structure to the 1+2 CFS of the $J=1$ level in the axial crystal environment. As discussed above, simulations indicate that the 10V and 8V the trapping sites predominate.

Second, it is well-known from experiments in atomic isolation and cooling in magnetic traps and their theoretical interpretation \cite{Hancox2004,Krems2005} that interaction anisotropy of the open-shell $4f^n6s^2$ lanthanide atoms with He is strongly suppressed. As both $4f$ and $5d$ shells of the Yb atom are beneath the outer spherical $6s^2$ shell, the same suppression should be expected for the states arising from the $4f^{13}5d^16s^2$ configuration. For the same reason, the interaction potentials of these states with RG atoms should be similar to those for the $4f^{14}6s^2$ ground state. This accounts for why the peaks in Figs.~\ref{fig:abs_fine} (c) to (e) are essentially diagonal Frank-Condon envelopes accompanied by phonon overtones. This also explains why the 1+2 crystal field splittings of the $4f \rightarrow 5d$ bands do not exceed 100 cm$^{-1}$, whereas they are $\sim$200 and $\sim$300 cm$^{-1}$ for the outer-shell $6s^2\ ^1\!S_0 \rightarrow 6s6p\ ^3\!P_{1}$ and $6s^2\ ^1\!S_0 \rightarrow 6s6p\ ^1\!P_{1}$ absorptions, respectively. 

Smaller $4f^{13}5d^16s^2$ state anisotropy should also diminish Jahn-Teller splitting of the doublet peaks. Observed splittings for bands (c), (d) and (e) are 25, 9 and 21 cm$^{-1}$, respectively. Significantly larger values of 80 and 140 cm$^{-1}$ for bands (a) and (b), respectively, are predicted by the $6s \rightarrow 6p$ absorption simulations. Similarity of the pairs of excited- and ground-state interaction potentials in  $4f \rightarrow 5d$ transitions to those in $6s \rightarrow 6p$ transitions implies similar frequencies for the phonons involved. Calculations for the 10V site simulated in Sec.\ref{Sec:spabsorption} predict 16 and 19 cm$^{-1}$ for $A_1$ and $E$ symmetry phonons, respectively---close to the value of 14 cm$^{-1}$ deduced from the sideband progression of $A_1$ symmetry phonons in Fig.~\ref{fig:abs_265}.


\subsection{Emission Spectrum}
\label{sec:demission}

The spectrum resulting from  $6s^2\ ^1\!S_0 \rightarrow 6s6p\ ^1\!P_1$ excitation has been discussed in detail in earlier reports \cite{xu11, Lambo_2012}. They found corroborating evidence for the presence of a crystal field in the observation of $6s6p\ ^3\!P_0$ state decay. The quenching of this state is posited to occur through a Stark-like coupling to the $6s6p\ ^1\!P_1$ and $6s6p\ ^3\!P_1$ states mediated by a field the order of $\sim$ 20 MV/m \cite{xu14}. A common feature of these reports, though, is that they only considered the fluorescence products at energies below those of the excitation energy i.e., in the $ < 2.6\times10^{4}$ cm$^{-1}$ range. The present work is the first one to consider the spectrum due to decay from higher energy levels, which are reached through the absorption of a photon by the excited Yb($^3\!P_0$) atom. Thus, the emission counterparts to the absorption bands (c), (d) and (e) in Fig. \ref{fig:abs_fine} are observed at 29,030 cm$^{-1}$, 37,730 cm$^{-1}$ and 39,610 cm$^{-1}$ in Fig. \ref{fig:emis_fine}. 

The above is important for the proposed single atom detection of Yb in sNe \cite{loseth19}. The imaging of a single $n$s$^2$ $S$ ground state atom in RG isolation has already been demonstrated for Ba in solid Xe through the observation of the 6s6p $^1\!P_1$ state decay following its resonant excitation \cite{chambers19}. But efficient detection of singlet state fluorescence is only possible when its branching ratio to other states is low. Thus, for the majority of M/RG systems, in which such fluorescence is heavily quenched by matrix-enhanced intersystem crossing, a proof-of-principle using the Yb/Ne system is highly relevant. The latter's $^3\!P_1$ state fluorescence has the advantage of being spectroscopically distinct from the source's exciting photons and its decay rate of $\sim$ 1.47 MHz is orders of magnitude greater than the dark count rate ($\sim$ 25 Hz) of widely-available single photon counting modules (e.g. Perkin Elmer AQR-16 SPCM). A drawback, though, of leveraging its triplet manifold for single atom detection is that its relatively long-lived $^3\!P_0$ state acts as a population trap into which electrons will be transferred on a timescale that is the order of that state's lifetime ($\sim$17 s in sNe). Fortunately, the observation here of the $4f^{13}5d^16s^2 \rightarrow 6s^2\ ^1\!S_0$ transition, whose decay rate is at least an order of magnitude higher than that of the $^3\!P_1$ state, shows that $6s6p\ ^3\!P_0 \rightarrow 6p^2\ ^3\!P_1$ repumping is a viable method for returning the dark state Yb atom back to the ground state.  

For metal atoms isolated in the heavier RGs, structures observed in the absorption spectra due to matrix effects such as multiple site occupation, Jahn-Teller coupling and CFS, usually have counterparts in the emission spectra. In sNe, however, this is rarely the case and emission bands typically have a simple unstructured Gaussian profile \cite{Healy2012}. Some matrix effects can still be resolved by lifetime measurements, as each component of the emission band may have distinct decay probabilities. For instance, previous lifetime measurements of the Yb($^3\!P_0$) state in sNe were fitted to a triple exponential and the various decay constants attributed to multiple site occupation and isotope composition \cite{xu11}. However, to identify such components in the lifetimes of all the excited state decays reported here would require a significant upgrade in our apparatus and so is left for future work. 

Spin-polarization of $^{171}$Yb nuclei in a matrix by optical pumping would enable the kinds of precision experiments envisioned for optically addressable solid-state spin systems. If the hyperfine structure of a particular transition is resolved, then it is expected that the optical pumping efficiency (the fraction of angular momentum transferred from the photon to the atom) will be high. However, if the hyperfine structure is unresolved, the optical pumping efficiency will likely depend on the details of the line-broadening mechanism. The hyperfine structure for the naturally abundant $^{171}$Yb($I=1/2$,14\%) and $^{173}$Yb($I=5/2$,16\%) isotopes do not appear to be resolved for any of the transitions studied here.
This implies that the optical pumping efficiency maybe low, which could be compensated for by a high number density. Because the transitions involving the inner shell $4f$ electrons have unusually narrow features, a more detailed study of them using circularly polarized light has the potential to find evidence of optical pumping.

\section{Conclusion}
\label{Sec:Summary}

High-resolution spectroscopy of Yb atoms in a cryogenic Ne matrix over a wide excitation range revealed five asymmetric absorption features. The two lowest are assigned to outer-shell $6s \rightarrow 6p$ transitions to $6s6p ^3P_1$ and $^1P_1$ atomic states and appear, respectively, as a broad doublet and an asymmetric triplet. Three bands that correspond to inner-shell $4f \rightarrow 5d$ transitions to distinct $J=1$ states arising from the $4f^{13}5d^16s^2$ configuration are much narrower and have rich structural features.

Interpretation of the band structures relied on the modeling of the stable trapping sites of the Yb atom in a perfect Ne fcc crystal. This indicated that the most energetically stable site is a 10-atom vacancy of $C_{4v}$ symmetry and that there are three other stable sites---6-, 8- and 13-atom vacancies---lying relatively close by in energy. In this environment, any $J=1$ atomic state should split into one-dimensional $A_1$ and two-dimensional $E$ representation. The dynamic Jahn-Teller effect will then lift the remaining degeneracy producing the generic 1+2 or 2+1 absorption bandshapes observed. The magnitude of the splittings depends on the interaction anisotropy which varies from strong for the $^1P_1$ state to medium for the $^3P_1$ state and is largely suppressed for the $4f^{13}5d^16s^2$ states. 

Similarly, the degree of phonon excitation reflects the difference in the ground- and excited-state interaction potentials. It ranges from significant band broadening in the $6s \rightarrow 6p$ transitions to the perfectly resolved phonon sideband progression in the $6s^2\ ^1\!S_0 \rightarrow 4f^{13}5d^16s^2\ (5/2,5/2)_1$ transition. While multiple site occupation complicates the detailed assignment of all features of the $4f \rightarrow 5d$ transitions, the evidence of resolved individual phonon lines for CFS components is very strong as their fits give plausible values for the Huang-Rhys parameter and the lattice displacement. 

Following the Mn/Kr \cite{by10} and Eu/Ar systems \cite{br11}, the Yb/Ne system is only the third example of a metal-rare gas matrix whose crystal field states yield ZPLs with accompanying phonon sidebands. It is the first one for which this occurs on an allowed transition, for which sNe is the matrix host and for which the trapping site has non-polyhedral symmetry. These results also indicate the peculiarities of the semi-quantum Ne matrix. On the one hand, well-defined CFS reflects the classical nature of relatively rigid trapping sites. On the other, the existence of multiple spacious trapping sites of axial symmetry and discernible individual phonon excitations manifest the onset of quantum solid-state physics. 


\section{Acknowledgment}

We would like to thank T. Oka and R. W. Dunford for helpful discussions and the use of their equipment. Experimental work and analysis is supported by Department of Energy, Office of Nuclear Physics, under Contract No. DEAC02-06CH11357, while the simulations are performed in the frame of Russian Science Foundation project No. 17-13-01466. H. X. and S. T. P. are supported by the U.S. Department of Energy, Office of Science, Office of Basic Energy Sciences, Division of Chemical Sciences, Geosciences, and Biosciences under contract No. DE-AC02-06CH11357. J. T. S. is supported by Argonne Director's postdoctoral fellowship and R. L. is supported by the China Postdoctoral Science Foundation. 


\end{document}


\begin{CJK*}{UTF8}{gbsn}

\preprint{draft version: \today}

\title{High Resolution Spectroscopy of Neutral Yb Atoms in a Solid Ne Matrix 
Supplemental Material}

\author{R. Lambo}
\thanks{These two authors contributed equally.} 
\affiliation{Shenzhen Institutes of Advanced Technology, Chinese Academy of Sciences, Shenzhen, 518055, China}
\author{C.-Y. Xu (徐晨昱)}
\thanks{These two authors contributed equally.} 
\affiliation{Physics Division, Argonne National Laboratory, Argonne, Illinois 60439, USA}
\affiliation{Department of Physics and Enrico Fermi Institute, University of Chicago, Chicago, Illinois 60637, USA}
\author{S. T. Pratt}
\affiliation{Chemical Science and Engineering Division, Argonne National Laboratory, Lemont, Illinois 60439, USA}
\author{H. Xu (徐红)}
\affiliation{Chemical Science and Engineering Division, Argonne National Laboratory, Lemont, Illinois 60439, USA}
\author{J. C. Zappala}
\affiliation{Physics Division, Argonne National Laboratory, Argonne, Illinois 60439, USA}
\author{K. G. Bailey}
\affiliation{Physics Division, Argonne National Laboratory, Argonne, Illinois 60439, USA}
\author{Z.-T. Lu (卢征天)}
\affiliation{Hefei National Laboratory for Physical Sciences at the Microscale, CAS Center for Excellence in Quantum Information and Quantum Physics, University of Science and Technology of China, 96 Jinzhai Road, Hefei 230026, China}
\author{P. Mueller}
\affiliation{Physics Division, Argonne National Laboratory, Argonne, Illinois 60439, USA}
\author{T. P. O'Connor}
\affiliation{Physics Division, Argonne National Laboratory, Argonne, Illinois 60439, USA}
\author{B. B. Kamorzin}
\affiliation{CEST, Skolkovo Institute of Science and Technology, Skolkovo Innovation Center, Moscow  121205,  Russia}
\author{D. S. Bezrukov}
\affiliation{CEST, Skolkovo Institute of Science and Technology, Skolkovo Innovation Center, Moscow  121205,  Russia}
\affiliation{Department of Chemistry, M.V. Lomonosov Moscow State University, Moscow 119991, Russia}
\author{Y. Xie}
\affiliation{Shenzhen Institutes of Advanced Technology, Chinese Academy of Sciences, Shenzhen, 518055, China}
\author{A. A. Buchachenko}
\thanks{Electronic address: a.buchachenko@skoltech.ru} 
\email{a.buchachenko@skoltech.ru}
\affiliation{CEST, Skolkovo Institute of Science and Technology, Skolkovo Innovation Center, Moscow  121205,  Russia}
\author{J. T. Singh}
\thanks{Electronic address: singhj@frib.msu.edu} 
\affiliation{Physics Division, Argonne National Laboratory, Argonne, Illinois 60439, USA}
\affiliation{National Superconducting Cyclotron Laboratory, Michigan State University, East Lansing, Michigan 48824, USA}

\begin{abstract}
Supplemental Materials describe the procedures and additional results of the modeling used to interpret experimental data from the Yb/Ne absorption spectroscopy.
\end{abstract}

\maketitle

\end{CJK*}



\section{Stable trapping sites}
\label{Sec:Sites}

\subsection{Matrix model}
The general approach for identifying thermodynamically stable crystal trapping sites of the matrix-isolated atoms and dimers is described in a few previous publications  \cite{Tao2015, Kleshchina2017, qwer, Tarakanova}. It considers the accommodation of the impurity atom in various vacancies of the bulk crystal and involves two steps, a search for the lowest energy structures that are stable in configuration space followed by a thermodynamic stability analysis by means of the convex hull concept. 

Computationally, the matrix was represented by a large spherical fragment of an ideal crystal containing a smaller concentric sphere. The smaller sphere encircled a movable set of host atoms, which can be removed or displaced to create a vacancy and should follow the potential field imposed by all other atoms during the minimization. The remaining host atoms kept their lattice positions to maintain a long-range crystal field. To investigate the number of host atoms that the guest atom can stably substitute, an $N$-atom vacancy was created near the center of the movable region and the guest atom inserted into it. Successive sampling of the parent structures for each $N$ followed by minimization (by the method of steepest descent) with the predefined force field gave the final structures possessing the lowest energy $E_\mathrm{min}$ as a function of $N$. The accommodation energy was defined as 
\begin{equation}
	E(N) = E_\mathrm{min}(N) - E_\mathrm{cryst} - NE^\mathrm{at},
\end{equation}
where $E_\mathrm{cryst}$ is the energy of the initial ideal crystal fragment and the last term accounts for the energy required to displace the atoms through $E^\mathrm{at}$, the crystal atomization energy per atom.

The dependence of the accomodation energy, $E(N)$, on $N$ identified the thermodynamically stable sites as those whose energies lie on its convex hull, in close analogy to the analysis of the discrete variable composition phase diagrams~\citenum{Zhu2013}.

For the present simulations, we took the Ne lattice to be a face-centered cubic (fcc). The radii of the crystal fragment and the movable atom region were set to $6a$ and $3a$, respectively, for which $a$ is the optimized fcc lattice parameter. Thus, approximately 3100 fixed and 460 movable Ne atoms were included in the simulations. The variation in the energies obtained with respect to system size was of only a few wavenumbers. The vacancies formed by the removal of $N=0,...20$ Ne atoms were analyzed.
 
\subsection{Force field}

To describe the host-host and guest-host interactions, the pairwise approximation was adopted. For the Ne--Ne interactions, we used the gas-phase potential derived by Aziz and Slaman \cite{Aziz1989} $V_\mathrm{gas}(r)$, where $r$ stands for the internuclear distance. It was further modified empirically to $V_\mathrm{NeNe}(r) = 0.7033\times V_\mathrm{gas}(1.0446\times r)$ to reproduce the  Ne fcc lattice parameter $a=4.464$ {\AA} and $E^\mathrm{at} = 165$ cm$^{-1}$, as described in Ref.~\citenum{Bezrukov2019}. The {\it ab initio} ground-state Yb($^1S_0$)--Ne potential was taken from Ref.~\citenum{YbNe2012}.

\subsection{Structure of the stable sites}

\begin{figure}[t]
\begin{center}
	\includegraphics[width=0.5\textwidth]{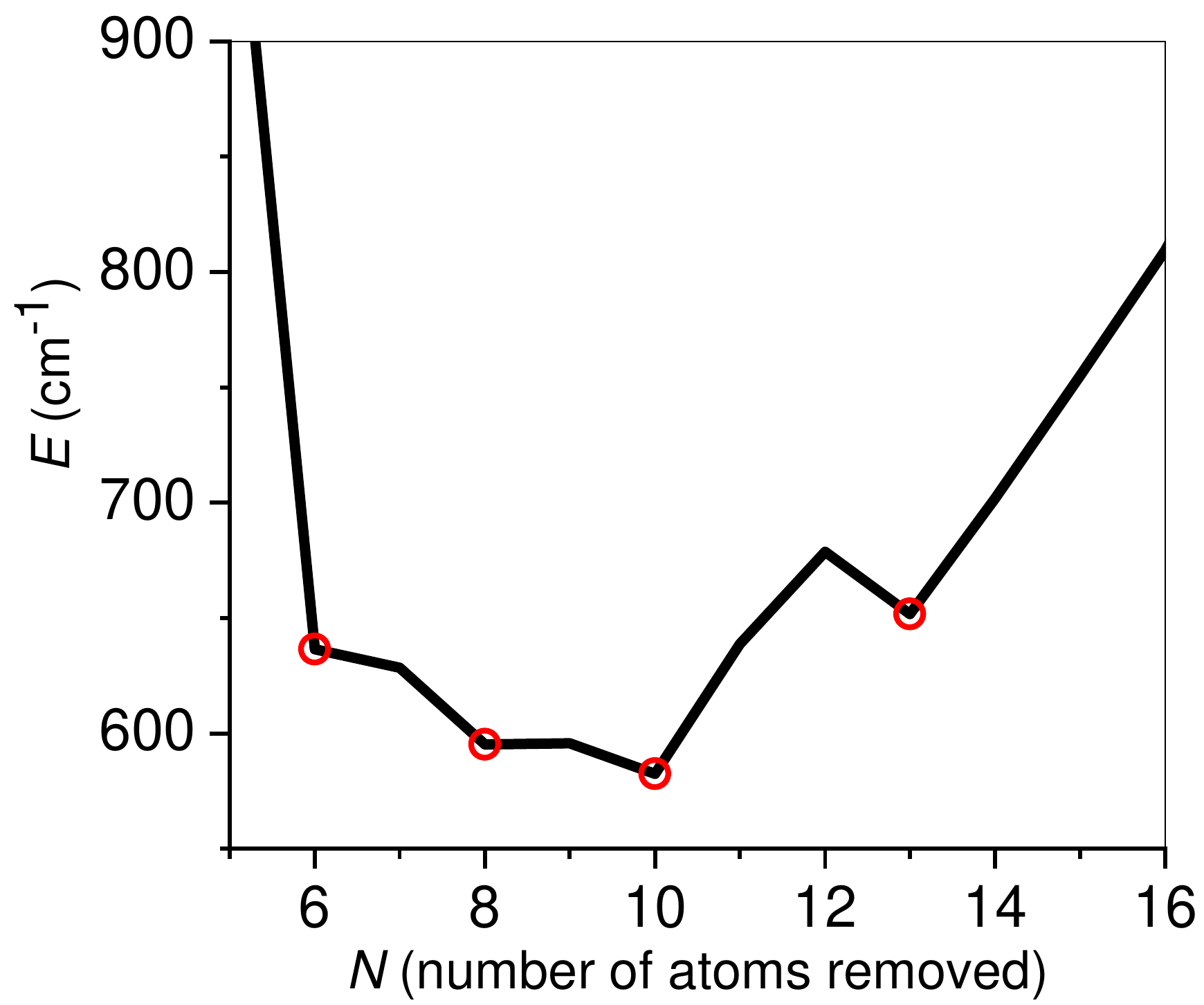}
\caption{Accommodation energy of the Yb atom in the Ne fcc crystal as a function of $N$, the number of Ne atoms removed from the lattice (solid line). Circles indicate the points lying on the convex hull of the $E(N)$ diagram. 
}\label{fig:convex}
\end{center}
\end{figure}

\begin{figure}[t]
\centering
\includegraphics[width=1\textwidth]{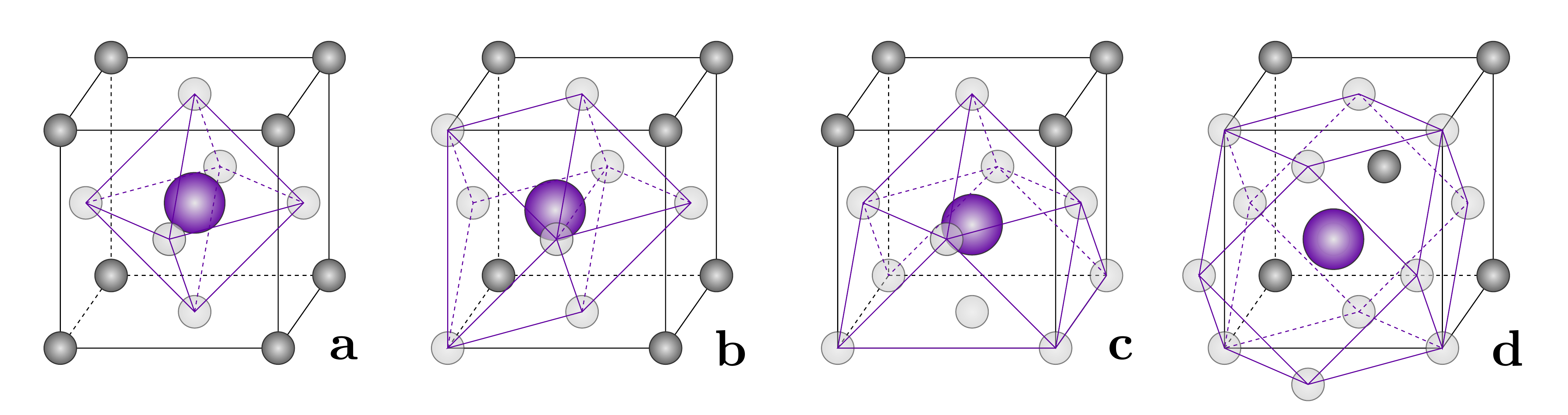}
\caption{Schematic structures of the Yb/Ne stable trapping sites, 6V (a), 8V (b), 10V (c) and 13V (d). Yb and Ne atoms are represented by large and small spheres, respectively; light grey spheres indicate the lattice positions of Ne atoms removed from the system; dark grey spheres indicate the lattice positions of the remaining Ne atoms. Note that for the 6V site Yb stays in the center of fcc octahedral void. In the axially-symmetric 8V and 10V sites---both of which can derived by further removing atoms from the 6V site---the Yb atom is somewhat displaced from the the void center. The perfect cuboctahedral site, 13V, originates from a single substitution accompanied by the removal of 12 nearest-neighbors. Image (d) corresponds to the substitution of all but one face of the fcc Ne atoms. }
\label{fig:struct}
\end{figure}

The resulting $E(N)$ diagram is shown in the Fig.~\ref{fig:convex}.  Convex hull analysis identified four thermodynamically stable sites for the Yb/Ne system, designated as 6V, 8V, 10V, and 13V for ``$N$-atom Vacancy''. Their structures are schematically shown in Fig.~\ref{fig:struct} with the reference to a single fcc unit cell.  Table~\ref{tab:geometry} characterizes the geometries of these sites in terms of the distance $R$ between the Yb atom and the nearest shell of surrounding Ne atoms. They are compared to the nearest-neighbor distance in the ideal Ne crystal and the equilibrium distance of the Yb($^1S_0$)--Ne complex. In all cases, even for $N=13$, Yb inflates the coordination shells with respect to the ideal crystal lattice by more that 30\%. The optimized Yb--Ne distances correspond to the repulsive region of the Yb($^1S_0$)--Ne potential resulting in positive accommodation energies for all sites. Table~\ref{tab:geometry} also specifies the point group describing the local symmetry of the Yb environment and the Einstein frequency of the Yb atom vibration in each site. 

\begin{table}[h!]
	\caption{$R$ is the distance between Yb and the nearest Ne atom (the number of atoms at the same distance is given in parentheses) for each stable trapping site. The nearest-neighbor Ne--Ne crystal distance and the Yb($^1S_0$)--Ne equilibrium distances are also given for comparison. The point symmetry group indicated describes the local environment of the Yb atom. The Einstein frequency $\omega$ of the Yb atom is given together with the vibrational mode degeneracy factor in parenthesis.} 
	\label{tab:geometry}
	\begin{center}
		\begin{tabular}{l@{\quad}c@{\quad}c@{\quad}c@{\quad}c@{\quad}c@{\quad}c@{\quad}}
			\hline\hline
			Site & $R$ ({\AA}) & Symmetry & $\omega$ (cm$^{-1}$)      \\
			\hline\hline
			6V   & 4.25(8) & $O_h$  & 15(3)  \\
			8V   & 4.28(4), 4.51(2), 4.57(2) & $C_{2v}$ & 10, 12, 15 \\
			10V  & 4.46(4), 4.59(4) & $C_{4v}$ & 10(2), 12 \\
			13V  & 4.59(6) & $O_h$    & 7(3)\\
			\hline
			Ne--Ne   & 3.16(12) & $O_h$ & --- \\
                   \hline
                   Yb--Ne  & 5.23 & $C_{\infty v}$ & 7.4 \\
			\hline\hline
		\end{tabular}
	\end{center}
\end{table}

\section{The $6s^2 \, {^1}S_0 \to 6s6p \, ^3P_1, \, ^1P_1$ absorption spectra}
\label{Sec:1P}

\subsection{Diatomics-in-molecule model}
\label{sec:DIM}

In the non-relativistic (NR) limit, both $^3P$ and $^1P$ states of atomic Yb posses triple degeneracy in the projection $\Lambda=\Sigma, \Pi$ of the orbital electronic angular momentum $L=1$. In the former case, it is multiplied by triple degeneracy in the projections of spin momentum $S$. The vectorial spin-orbit (SO) interaction couples $L$ and $S$ producing three distinct atomic multiplets, or fine-structure levels, with total electronic momentum $J = 0, 1$ and 2 and $2J+1$ degeneracy. The $^1P_1$ SO state is equivalent to a $^1P$ state that can be subjected only to second-order SO interactions with other atomic terms in the presence of interatomic interactions, such as the ground state $^1S_0$. 

Depending on the strengths of atomic SO coupling and atom-matrix interactions, one can consider strong- and weak-field limits. In the former limit, the matrix problem should be solved non-relativistically and SO coupling added afterwards. The latter limit, which implies the solution of the matrix problem for each SO multiplet separately, is applicable to the $^3P_1$ term in the present case. Indeed, the strength of the Yb-Ne interaction does not exceed 30 cm$^{-1}$ leading to a matrix-induced shift in the excitation frequency, $\Delta \sigma$, of about 500 cm$^{-1}$ and a difference in the stable trapping site energies of less than 100 cm$^{-1}$. By contrast, the $^3P_J$ levels span 2500 cm$^{-1}$ and its SO coupling constant is greater than 800 cm$^{-1}$. 

The weak coupling picture allowed us to describe the $^3P_1$ and $^1P_1$ states on the same footing using the well-know diatomics-in-molecule model for the global potential energy surfaces (PESs) arising from the NR atomic $P$-state \cite{Boatz1994,Visticot1994,Krylov1996,Ryan2010}. It requires only straightforward replacement of the Yb($^{2S+1}P$)-Ne potentials for $\Sigma$ and $\Pi$ components by the Yb($^{2S+1}P_J$)--Ne $\Omega = 0$ and 1 components, respectively; or, in more conventional notation, the replacement of $V_\Sigma$ and $V_\Pi$ by $V_0$ and $V_1$, respectively. Note that to obtain the $\Omega$ potentials, vectorial SO coupling has to be included explicitly for the Yb--Ne pair. 

In brief, for each Yb--Ne pair three PESs correlating to the $^{2S+1}P_1$ Yb level were described by a diagonal matrix in the molecular-fixed frame,
\begin{equation*}
	H_{\text{mf}} = 
	\begin{pmatrix}
		V_0 & 0     & 0     \\
		0        & V_1 & 0     \\
		0        & 0     & V_1 \\
	\end{pmatrix}
	,
\end{equation*}
which, after the angular transformation to the space-fixed frame, acquired the form
\begin{equation*}
	H_{\text{sf}} = 
	\begin{pmatrix}
		-\Delta\cos^2\theta & \frac{1}{2}\Delta\sin2\theta \sin\phi & -\frac{1}{2}\Delta\sin2\theta \cos\phi \\
		\Delta\frac{1}{2}\sin2\theta \sin\phi & -\Delta\sin^2\theta \sin^2\phi+V_1 & \frac{1}{2}\Delta\sin^2\theta \sin2\phi \\
		-\frac{1}{2}\Delta\sin2\theta \cos\phi & \frac{1}{2}\Delta\sin^2\theta \sin2\phi & -\Delta\sin^2\theta \cos^2\phi+V_1
	\end{pmatrix}
	,
\end{equation*}
where $\Delta = (V_1 - V_0)$, and $\phi, \theta$ are the Euler angles for the particular Yb--Ne molecular reference. Summing up the matrices for all Ne atoms in the system, extracting redundant atomic excitation energies and adding all pairwise Ne-Ne interactions, we arrived at the final matrix whose eigenvalues define the three desired PESs. 

The {\it ab initio} SO-coupled Yb--Ne potentials for $^3P_1$ and $^1P_1$ multiplets were taken from Ref.~\citenum{YbNe2012}. It should noted that they include both first- and second-order SO couplings in the whole manifold of states correlating to the Yb atom in the $6s^2$, $6s6p$ and $6s5d$ electronic configurations. 

\subsection{Absorption frequencies and band shapes}
\label{sec:abs}

Vertical transition frequencies were calculated for the equilibrium geometries of the trapping sites using the same matrix model. 

To compute the band shapes, the thermodynamic integration approach, similar to that described in Ref.~\citenum{Ryan2010}, was employed. It consisted of a molecular dynamic simulation of the effect of the system's evolution in time on the ground-state PES. At each simulation step, the instantaneous vertical transition energies to all three excited PESs were calculated to obtained the histograms representing partial spectral envelopes. Their sum gave the total spectral envelope.  

In particular, we used Langevin dynamics, simulating the NVT ensemble at effective temperature $T_\mathrm{eff}$ that classically matches the quantum vibrational probability of a system at real temperature $T$ \cite{Ryan2010}: 
\begin{equation*}
T_\mathrm{eff}  = \frac{\hbar\omega_0}{2k_\mathrm{B}}\left(\tanh{\frac{\hbar\omega_0}{2k_\mathrm{B}T}}\right)^{-1},
\end{equation*}
where $k_\mathrm{B}$ is the Boltzmann constant and $\omega_0$ is the matrix phonon frequency. For the Ne matrix at $T=5$ K, $T_\mathrm{eff} = 34$ K. The friction coefficient for the Langevin thermostat was taken as $\gamma=5$, a value which provided good energy conservation upon switching the thermostat on and off. 

Spectral simulations were performed for a smaller matrix system than that used in the search for stable trapping sites. The radii of the crystal fragment and the movable atom region were set to $4a$ and $2.5a$, respectively, so that approximately 800 fixed and 230 movable atoms were included in the simulations. The variation in the energies obtained with respect to the system size is a few tens of wavenumbers.   

\begin{figure}[h]
\centering
\includegraphics[width=0.8\textwidth]{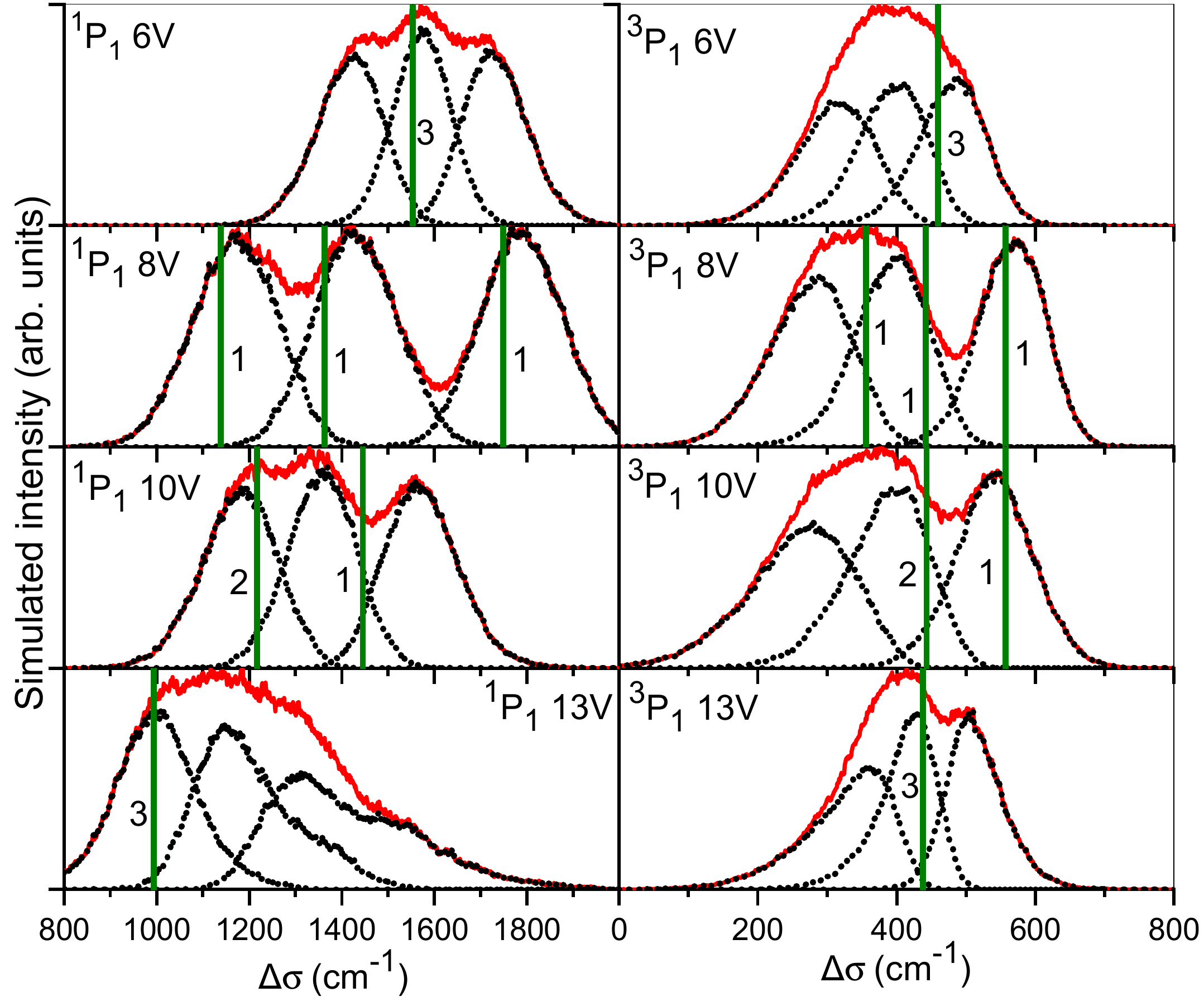}
\caption{Simulated absorption spectra of the Yb  $6s^2 \, {^1}S_0 \to 6s6p \, ^1P_1$ (right panels) and $6s^2 \, {^1}S_0 \to 6s6p \, ^3P_1$ (left panels) transitions in Ne. The spectra are simulated for 6V, 8V, 10V and 13V sites (from top to bottom). In addition to the total envelopes shown by solid lines, partial envelopes for each of three excited PESs are shown by dotted lines. Vertical transition energies are indicated by the vertical bars with the degeneracy indicated. They mark transitions to certain crystal field components. Further splitting is caused by the Jahn-Teller effect. }
\label{fig:bands}
\end{figure}

Figure~\ref{fig:bands} summarizes the main results. Overall, the simulated bandshapes accord nicely with the symmetry of the trapping sites, which dictates the crystal field and Jahn-Teller splittings. The bands of the $6s^2 \, {^1}S_0 \to 6s6p \, ^1P_1$ transition are more than twice as wide as those of the $6s^2 \, {^1}S_0 \to 6s6p \, ^3P_1$ transition. This is due to larger crystal field splittings (marked by the vertical lines in Fig.~\ref{fig:bands}) and the Jahn-Teller effect, which lifts the remaining degeneracy. In the case of the triplet $P$-state, individual Jahn-Teller components are not resolved. Both effects reflect the significantly larger anisotropy of the Yb--Ne interaction in the singlet $P$-state. In addition, the triplet state interaction potentials are much closer to the ground-state potential, which results in very similar vertical transitions for all trapping sites. The peculiar asymmetric shapes obtained for the perfectly octahedral 13V site can be explained by the large dynamical distortion of its loose and labile structure.